\documentclass[5p]{elsarticle} %
\usepackage{verbatim}
\usepackage{color}
\journal{Computational Materials Science}

\usepackage{float}
\usepackage[export]{adjustbox}
\usepackage[]{subfigure}
\usepackage{placeins}
\usepackage[english]{babel}  
\usepackage[utf8x]{inputenc}
\usepackage{graphicx}
\usepackage{booktabs}
\usepackage{tabularx}
\usepackage{caption}
\usepackage{amsmath}
\usepackage{amssymb}
\usepackage{multirow}

\begin{document}
\begin{frontmatter}
\title{A Comparison between Monte Carlo Method and the Numerical Solution of the Ambartsumian-Chandrasekhar Equations to Unravel the Dielectric Response of Metals}

\author[ECT,DICAM]{Martina Azzolini}

\author[Russia,Wien]{Olga Yu. Ridzel}

\author[Russia]{Pavel S. Kaplya}

\author[Russia]{Viktor Afanas'ev}

\author[DICAM,QUEEN]{Nicola M. Pugno}

\author[ECT]{Simone Taioli\corref{mycorrespondingauthor1}}
\cortext[mycorrespondingauthor1]{Corresponding author}
\ead{taioli@ectstar.eu}
\author[ECT]{Maurizio Dapor\corref{mycorrespondingauthor2}}
\cortext[mycorrespondingauthor2]{Corresponding author}
\ead{dapor@ectstar.eu}

\address[ECT]{European Centre for Theoretical Studies in Nuclear Physics and Related Areas (ECT*-FBK) and Trento Institute for Fundamental Physics and Applications (TIFPA-INFN), Trento, Italy}
\address[Wien]{Institute of Applied Physics, Vienna University of Technology, Vienna, Austria}
\address[Russia]{Department of General Physics and Nuclear Fusion,  National Research University ``Moscow Power Engineering Institute'', Moscow,
Russia}
\address[DICAM]{Laboratory of Bio-Inspired and Graphene Nanomechanics - Department of Civil, Environmental and Mechanical Engineering, University of Trento, Italy}
\address[QUEEN]{School of Engineering and Materials Science, Materials Research Institute, Queen Mary University of London, UK}
%\address[ASI]{Ket Lab, Edoardo Amaldi Foundation, Italy}

\begin{abstract}
In this work we describe two different models for interpreting and predicting Reflection Electron Energy Loss (REEL) spectra and we present results of a study on metallic systems comparing the computational cost and the accuracy of these techniques. These approaches are the Monte Carlo (MC) method and the Numerical Solution (NS) of the Ambartsumian-Chandrasekhr equations. The former is based on a statistical algorithm to sample the electron trajectories within the target material for describing the electron transport.  
The latter relies on the numerical solution of the Ambartsumian-Chandrasekhar equations using the invariant embedding method. Both methods receive the same input parameters to deal with the elastic and inelastic electron scattering. 
To test their respective capability to describe REEL experimental spectra, we use copper, silver, and gold as case studies. Our simulations include both bulk and surface plasmon contributions to the energy loss spectrum by using the effective electron energy loss functions and the relevant extensions to finite momenta. The agreement between MC and NS theoretical spectra with experimental data is remarkably good. 
Nevertheless, while we find that these approaches are comparable in accuracy, the computational cost of NS is several orders of magnitude lower than the widely used MC. \\Inputs, routines and data are enclosed with this manuscript via the Mendeley database. 
\end{abstract}
%\vspace{2pc}

\end{frontmatter}
\onecolumn

\section{Introduction}
\label{intro}
Electron beam analysis is a widely used tool for materials characterization \citep{hillier1944microanalysis} due to the ease of handling, detecting, bending, and counting charges with high spatial and spectral resolution by electromagnetic fields. In particular, Reflection Electron Energy Loss (REEL) spectroscopy enables one to evaluate both optical properties \citep{yubero1998optical} and chemical composition  \citep{nikzad1992quantitative, da2014monte} by probing the material under investigation via mono-energetic electron beams. \\
Specifically, after acceleration to the desired kinetic energy the electron beam ({\it primary beam}) interacts with the specimen through elastic and inelastic scattering processes. 
The fraction of primary electrons elastically reflected from the target surface appears in the spectrum as a peak of high intensity with respect to the inelastic background.  Typically, the inelastic background structure emerges from single-particle excitations, such as inter- and intra-band transitions, inner-shell and Auger decay ionisation, and collective excitations, such as phonon and plasmon oscillations.
Additionally, primary and secondary electrons, which are not trapped within the sample, emerge with characteristic energies that are the fingerprint of the underlying electronic structure of the solid.\\
\indent Thus, the materials response to the interaction with external electromagnetic fields can be used to access information on a variety of physical properties, particularly electronic and optical observables. Notably, this information is entirely encoded in the dielectric function of the material $\epsilon(q,W)$, which generally depends on both transferred momentum ($q$) and energy loss ($W$). With regard to charge transport within a medium, the knowledge of this quantity allows one to calculate the electron inelastic mean free path (IMFP) via the Ritchie theory \cite{RitchieHowie}. In the latter approach the IMFP depends on the energy-loss function (ELF), which is defined by the inverse of the imaginary part of the dielectric function. The ELF, which is only a property of the target material, is thus a fundamental quantity directly related to the energy deposited by charged particles \cite{egerton2011electron,TAIOLI2015191,PhysRevB.79.085432}. Typically, the ELF can be computed by using three different approaches \cite{doi:10.1002/sia.5947,doi:10.1002/sia.5878}: i) from ab initio simulations \cite{umari2012communication,taioli2009electronic}; ii) semi-empirically \cite{azzolini2017monte}, by using experimental measurements of optical or electron energy loss spectra (EELS); iii) from model calculations within the electron gas theory \cite{Kyriakou}. \\
\indent In practice, the accurate computation of the ELF over the whole momentum dispersion, is still challenging for ab-initio simulations due to the high computational cost of including local field effects (LFE), and exchange and correlation beyond the Random Phase Approximation (RPA). The dielectric function can be obtained by using either many-body perturbation theory (MBPT) \cite{umari2012communication,taioli2009electronic} or time-dependent density functional theory (TDDFT) \cite{doi:10.1002/sia.5878,azzolini2017monte,segatta2017quantum}. \\
\indent In the semi-empirical Drude--Lorentz (DL) model \cite{azzolini2018secondary,azzolini2018anisotropic}, which will be outlined below, the assessment of the ELF relies on experimental data, which typically cover only a limited range of the target excitation spectrum for vanishing momentum transfer (long wavelength limit of the dielectric response). Basically, within this model the experimental ELF at $q=0$ is equalized to the analytical ELF obtained by using the Lindhard dielectric function model in the plasmon-pole approximation \cite{Lindhard}. The extension to finite $q$, which accounts for the dispersion of the excitation spectrum along the momentum axis, is finally obtained by imposing typically a quadratic polynomial dependence on the transfer momentum modulus. This model works well for 3D metals, where the Fermi electron-gas theory can be safely applied. In the case of semi-infinite medium, such as a slab, one adds a linear dependence of the plasmon frequency on momentum transfer to deal with the surface plasmon dispersion \cite{doi:10.1002/sia.5878}. A variant to the DL approach is the Penn model \cite{PhysRevB.35.482}, where the ELF is written as a convolution of the imaginary part of the inverse Lindhard dielectric function with a spectral density function determined from the experimental optical data.\\ \indent Finally, a third approach is based on the use of the Mermin energy loss function (MELF), which provides automatically the ELF at finite transfer momenta starting from optical data through the analytic properties of the Mermin dielectric function \cite{PhysRevB.1.2362} without the need to make a choice on a particular dispersion relation. This is a fundamental difference with the previous cases. \\
\indent In this respect, we notice that the major shortcomings of the DL and MELF methods are represented by the use of i) experimental optical data, often showing substantial discrepancies among different data sets; ii) the RPA to the electron gas theory to treat the electron-electron interaction; iii) the first Born approximation, which neglects exchange effects and is strictly valid only at high energies or for small scattering potential. The latter are all sources of error in the assessment of the ELF via the DL and MELF models, which must be carefully evaluated. At variance, the ab-initio approaches, while computationally more expensive than the DL and MELF semi-empirical methods, can in principle deliver the most accurate results concerning the ELF \cite{azzolini2017monte}.\\
\indent In this work, in particular, we present two different computational methods for simulating REEL spectra: i) the Monte Carlo (MC) approach \citep{dapor2014transport,1749-4699-2-1-015002}, and ii) the numerical solution (NS) of the Ambartsumian-Chandrasekhar (AC) equations using the Invariant Embedding Method (IEM) \citep{afanas2016analytical}. These methods are applied to the simulation of the REEL spectra of three metals: copper, silver and gold. In order to compare the accuracy of these approaches, tests were performed by using the same input data with respect to elastic and inelastic interactions. In particular, we take into account both bulk and surface plasmons, for the latter by means of an effective dielectric function of the materials.  
Elastic interactions are assessed via the Mott theory \citep{mott1929scattering}, while energy losses due to inelastic scattering events are modelled via the Ritchie dielectric formalism \citep{Ritchie_PhysRev_1957}.\\ 
\indent Calculated data of elastic and inelastic cross sections, mean free paths, and cumulative probabilities \citep{database_input} along with the computer codes and routines used to perform these simulations~\citep{ inelastic_calc, REEL_NS_calc, elastic_calc}
have been made freely available in the Mendeley open access database.

\section{Computational methods}

\subsection{Monte Carlo approach}

In MC simulations, we assume that a mono-energetic electron beam impinges orthogonal to a target surface. The scattering centers within the material dissipate the primary electron beam via elastic and inelastic interactions. Within the MC approach one follows the trajectories of electrons during their entire way inside and outside the solid target, where they can be collected and recorded as a function of kinetic energy and/or emission angle.
The electron path is typically described by an exponential law, so that the step length ($\Delta s$) between two subsequent collisions is given by:
\begin{equation}\label{mfp}
\Delta s = - \lambda_\mathrm{tot} \ln(r), 
\end{equation}
where $\lambda_\mathrm{tot} = \left({1}/{\lambda_{\mathrm{el}}} + {1}/{\lambda_{\mathrm{inel}}}\right)^{-1}$ is the total mean free path, and $r$ is a random number uniformly distributed in the range 0 to 1. Otherwise specified, by $r$ we mean a different random number for each relevant sampling.\\
\indent The MC procedure applied to electron transport within the solid proceeds then in the following way: the probability to undergo inelastic collisions, given by $p_\mathrm{inel} = {\lambda_\mathrm{tot}}/{\lambda_{\mathrm{inel}}}$,  is compared with $r$. Should the condition $r < p_\mathrm{inel}$ be satisfied then the collision is classified as inelastic, otherwise is elastic. 
Upon inelastic events the electrons lose their kinetic energies according to the cumulative probability distribution:
\begin{equation}
\label{inelcum}
\mathrm{P}_\mathrm{inel}(E,W) = \lambda_\mathrm{inel} \int\limits_{0}^{W} \frac{d \lambda_\mathrm{inel}^{-1}}{dW'}dW', 
\end{equation}
which is a function of the kinetic energy $E$ and of the energy loss $W$.
$W$ is determined by generating a second uniformly-distributed random number $r$ in the range 0 to 1, and by finding the value of $\mathrm{P}_\mathrm{inel}$ that equalizes $r$. The determination of the directional change upon inelastic collision of the electron trajectories will be discussed below. \\ 
\indent At variance, elastically scattered electrons undergo only  directional change, which can be obtained by using the following elastic cumulative probability:
\begin{equation}
    \label{elcum}
    \mathrm{P}_\mathrm{el}(E, {\overline{\theta}}) = 2\pi\lambda_\mathrm{el} \int\limits_0^{\overline{\theta}}  \frac{d\lambda_\mathrm{el}^{-1}}{d\Omega}\sin{\vartheta}d\vartheta,
\end{equation}
which is determined for a given kinetic energy $E$ by varying the scattering angle $\theta$ in the range $[0,\overline{\theta]}$. 
The angular deflection of the trajectory is assessed by generating a third random number $r$, and by finding the value of the upper extreme of integration $\overline{\theta}$ in Eq. (\ref{elcum}) that equalizes $r$.
After a series of elastic and inelastic interactions, the electron can reach the target surface and can be released provided that the emission condition is fulfilled. This emission condition reflects the fact that the target-vacuum interface represents by all means an energy barrier that the electron has to overcome at the interface, and reads: 
\begin{equation}
E \cos^2 \alpha \ge \chi,
\label{eq:emissione}
\end{equation}
where $\alpha$ is the incident angle formed by the electron direction of motion inside the target material with respect to the surface normal, $E$ is the electron kinetic energy, and $\chi$ is the electron affinity or work function of the metals. To determine the secondary electron emission spectral features one needs to set the latter quantity for the materials under investigation. The values $\chi=5.4$ eV for copper (the experimental value is equal to 4.6 eV \cite{doi:10.1063/1.5000118}), $\chi= 4.4$ eV for silver (experimental value is 4.4 eV \cite{doi:10.1063/1.5000118}), and $\chi= 4.7$ eV for gold (experimental value is 5.3 eV \cite{doi:10.1063/1.5000118}) were set so to obtain the best agreement between MC simulations and the secondary electron experimental data of the metals \cite{azzolini2018secondary}. However, as we are interested in simulating high-energy REEL spectra (~1 keV) we do not assess here the secondary electron emission. \\ 
\indent A detailed description of how these quantities are computed will be presented in the sections \ref{ela} and \ref{inela}, while further details on our MC approach can be found in Refs. \citep{TAIOLI2010237, dapor2014transport}. 
The ensemble of trajectories used in MC simulations of REEL spectra is assessed so to reach statistical significance and low noise of the simulated data ($\approx 10^9$). 

\subsection{Numerical Solution (NS)}

\begin{figure}[h!]
    \centering
    \includegraphics[width = 0.55 \textwidth, angle = 0]{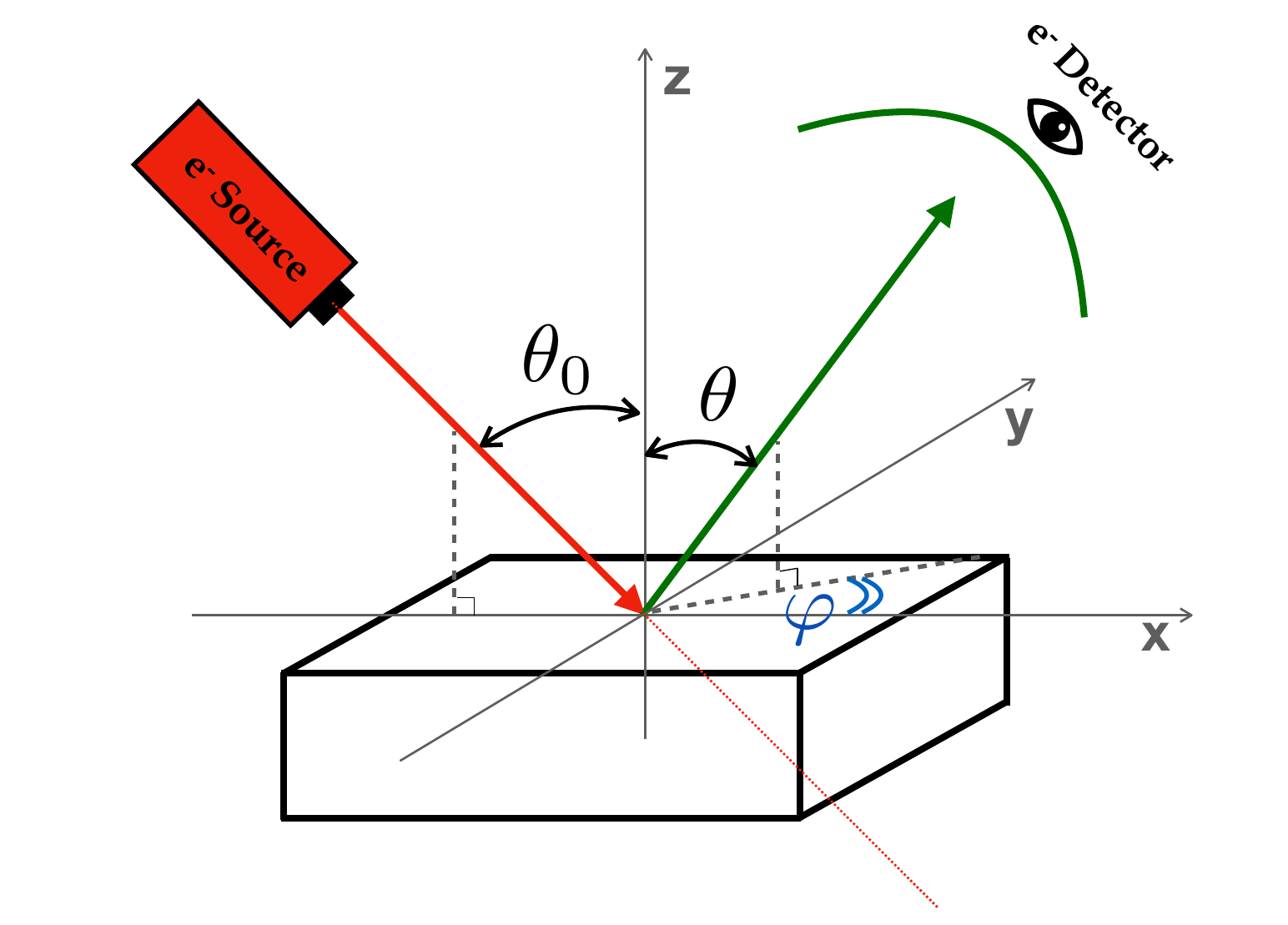}
    \caption{Experimental layout of a typical REELS experiment.}
    \label{fig:scattexp}
\end{figure}

The NS approach is based on the invariant embedding method developed by Ambartsumian and Chandrasekhar for radiative transfer \citep{chandras}. This method can be also applied to a system of non-linear equations to find the solution of the boundary problem for the electron transfer \citep{Afanas2017}. First, we introduce for the sake of clearness the parameters that are needed in the description of this method.\\
\indent The initial incident polar angle with respect to the surface normal is indicated by $\theta_0$, while the emission polar angle is named $\theta$ (see Fig. \ref{fig:scattexp} for a REELS typical experimental layout). The cosine of the incident angle is designated $\mu_0 = \cos(\theta_0)$, while the cosine of the emission angle is labelled $\mu = \cos(\theta)$. The azimuthal angle is denoted by $\varphi$. 
The transfer equation for the flux density $N(z,W,\vec{\Omega})$ of electrons travelling at depth $z$ in the direction $\vec{\Omega}=(\mu, ~ \varphi)$ with energy $E=E_0-W$ can be written as follows \citep{Werner2001}:
\begin{eqnarray}
\label{trasf} 
\mu \dfrac{\partial }{\partial z} N(z,W,\vec{\Omega})&=&-\frac{1}{\lambda_\mathrm{tot}}N(z,W,\vec{\Omega})+ \nonumber\\&+&\frac{1}{\lambda_\mathrm{inel}} \int\limits_0^W N(z,\varepsilon,\vec{\Omega})x_\mathrm{inel}(E_0-\varepsilon,W-\varepsilon) d\varepsilon+ \nonumber \\&+&\frac{1}{\lambda_\mathrm{el}}\int\limits_0^{2\pi}\int\limits_{-1}^{1} N(z,W,\vec{\Omega}') x_\mathrm{el}(E_0-W,\vec{\Omega}'\rightarrow\vec{\Omega})d\vec{\Omega}',
\end{eqnarray}
with the following boundary conditions: 
\begin{equation}
\left\{  \begin{matrix} N(0,0, \vec{\Omega}) &=& N_0 \delta(W)\delta(\vec{\Omega} -\vec{\Omega_0})\ & \mbox{if} \ 0 \le \mu \le 1,  \\ 
N(d,0, \vec{\Omega})  &=& 0  & \mbox{if} \ -1 \le \mu \le 0  , \end{matrix} \right.
\end{equation} 
where $E_0$ is the incident energy, $\vec{\Omega}_0=(\mu_0,0)$ is the initial direction, $d$ is the layer thickness, $x_\mathrm{inel}$ is the normalized probability distribution of the energy loss in a single inelastic event (also referred to as normalized differential inverse inelastic mean free path (NDIIMFP)): 
\begin{equation}
x_\mathrm{inel}(E, W)=\frac{d \lambda_\mathrm{inel}^{-1}(E, W)}{dW} \lambda_\mathrm{inel}
\end{equation}
which satisfies the normalization condition:
\begin{equation}
\label{eq:x_in_norm}
\int\limits_0^{E} x_\mathrm{inel}\left( E,W \right) dW = 1.
\end{equation}
In Eq. (\ref{trasf}) $x_\mathrm{el}(E,\vec{\Omega}'\rightarrow\vec{\Omega})$ is the normalized differential elastic scattering cross section (DECS) \citep{Salvat2005}. DECS must fulfill the following normalization condition:
\begin{equation}
\int\limits_0^{2\pi} \int\limits_{-1}^{1} x_\mathrm{el}(E,\vec{\Omega}) d\vec{\Omega}  = 2\pi.
\end{equation}

Eq. (\ref{trasf}) is solved within the one-speed approximation, so all the cross sections are fixed at the initial energy of the electron and do not change while it slows down. This means that $E$ in the NDIIMFP and DECS is replaced by $E_0$. Hereafter, we drop the explicit dependence on the $E_0$ variable of the cross sections for simplicity.

For the interpretation of REELS experiments one needs to estimate the reflected electron flux, described by the so-called reflection function $R\left(\tau,E_0,W,\vec{\Omega_0},\vec{\Omega}\right)$ as the ratio of the outgoing electron flux to the incoming one. Here and after the normalized depth $\tau=z/\lambda_{tot}$ is used. The reflection function within the Partial Intensity Approach (PIA) \citep{Afanasev1994,Werner2001} can be written as follows:
\begin{equation}
\label{eq:R_series}
R\left(\tau,E_0,W,\vec{\Omega_0},\vec{\Omega}\right)=  \sum\limits_{k=0}^\infty  R_{k}\left(\tau,E_0,\vec{\Omega_0},\vec{\Omega}\right)x_\mathrm{inel}^{k}\left(W\right)
\end{equation}
where the $x_\mathrm{inel}^k(W)$  is the $k$-fold self-convolution of the 
NDIIMFP or the energy loss spectrum after $k$ successive inelastic scattering events: 
\begin{equation}
\label{eq:x_in_k}
\begin{array}{l}
x_\mathrm{inel}^{k}\left( W \right) =\int\limits_0^W x_\mathrm{inel}\left( W -\varepsilon \right) x_\mathrm{inel}^{k-1}\left( \varepsilon \right) {d\varepsilon},
\end{array}
\end{equation}
In Eq. (\ref{eq:x_in_k})
$x_\mathrm{inel}^0(E_0, W)=\delta(W)$ is the Dirac function, and $x_\mathrm{inel}^1(E_0, W)=x_\mathrm{inel}(E_0, W)$ is the normalized probability distribution of the energy loss in a single inelastic event (Eq. (\ref{eq:x_in_norm})). By using $\mu$ and $\varphi$ to define the electron direction, 
$R_k\left(\tau,E_0,\mu_0,\mu,\varphi\right)$ can be rewritten most conveniently by expanding in azimuthal harmonics identified by the azimuthal index $m$:
\begin{equation}
R_k\left(\tau,E_0,\mu_0,\mu,\varphi\right) = \sum_{m = 0}^\infty (2 - \delta_{m0}) R^m_k\left(\tau,E_0,\mu_0,\mu\right)\cos(m \varphi),
\end{equation}
where $\delta_{mm'}$ is the Kronecker symbol.

The partial intensities $R^m_k\left(\tau,E_0,\mu_0,\mu\right)$ can be found by using the invariant embedding method (IEM) \citep{afanas2016analytical,Afanas2017}, which involves 
the following steps:
\begin{enumerate}
 \item add a layer to the bulk, which is thin enough to allow only one scattering event;
 \item consider single scattering processes in that layer, which contribute to the change of the reflection function of the system;
 \item find a solution for $R^m_k$ of the obtained system of equations.
\end{enumerate}
Following these steps the partial intensity coefficients $R^m_k\left(\tau,E_0,\mu_0,\mu\right)$ can be obtained as follows \citep{afanas2016analytical,Afanas2017}:
\begin{eqnarray}
\label{eq:dR_k}
\dfrac{\partial }{\partial \tau} R^m_k  + \left( \dfrac{1}{\mu _0} + \dfrac{1}{\mu } \right) R^m_k  &=& \Lambda x_\mathrm{el}^- + \Lambda R^m_k \ast x_\mathrm{el}^+ + \Lambda x_\mathrm{el}^+ \ast R^m_k + \Lambda R^m_0 \ast x_\mathrm{el}^-\ast R^m_k + \nonumber \\ &+& \left( 1-\delta _{0k}\right)\Lambda R^m_k \ast x_\mathrm{el}^- \ast R^m_0+ \Lambda \sum\limits_{j=1}^{k-1} {R^m_{k-j} \ast x_\mathrm{el}^- \ast R^m_j }  + \nonumber \\
&+& \left( 1-\Lambda \right) \left( \frac{1}{\mu _0} + \frac{1}{\mu } \right) R^m_{k-1},
\end{eqnarray}
where $\Lambda=\sigma_{el}/(\sigma_{el}+\sigma_{inel})$ is the single scattering albedo. By writing $R^m_k=R^m_k\left(\tau,E_0,\mu_0,\mu\right)$ we dropped the variable dependence to simplify the notation and the convolution operator $\ast$ is defined as follows:
\begin{equation}
\label{eq:conv}
F_1 \ast F_2 = \int\limits_{0}^{1} F_1\left(\mu_0,\mu'\right)
\cdot F_2 \left(\mu',\mu \right) \frac{d\mu'}{\mu'}.
\end{equation}
The '$+$' and '$-$' superscripts appearing in the DECS in Eq. (\ref{eq:dR_k}) refers to the sign of the cosine of the polar angle:
\begin{equation}
\label{eq:x+-}
\begin{array}{l}
\begin{cases}
x_\mathrm{el}^{+}\left(\mu_{0},\mu \right)=x_\mathrm{el}\left(\mu_{0},\mu \right), & \mathrm{sign}\left(\mu_{0}\cdot\mu\right)=1,\\
x_\mathrm{el}^{-}\left(\mu_{0},\mu \right)=x_\mathrm{el}\left(\pm\mu_{0},\mp\mu \right), & \mathrm{sign}\left(\mu_{0}\cdot\mu\right)=-1.
\end{cases}
\end{array}
\end{equation}
Here the '$-$' index is applied when the flux is backscattered, while the '$+$' sign is used if the flux direction is not reversed. Eq. (\ref{eq:dR_k}) implies the following boundary conditions:
\begin{equation}
\label{eq:bond_R}
R^m_k\left(0,E_0,\mu_0,\mu\right)=0,\> R^m_{k<0}=0.
\end{equation}
Eq. (\ref{eq:dR_k}) can be solved by discretizing the angular domain using e.g. the Gaussian quadrature points and weights.
Within this approach, we thus finally obtain differential matrix equations that can be solved by using either the backward differentiation formula (BDF) or the matrix exponential formalism \citep{Afanas2017}. The latter numerical method is based on the discrete ordinate formalism, which reduces the Ambartsumian-Chandrasekhar equations to the algebraic Riccati and Lyapunov equations \citep{afanas2016analytical}.\\ 
\indent Further details on the numerical calculations using this method can be found in Refs.~\citep{Afanasev2015,afanas2016analytical,Afanas2017}. The ESCal software based on the MATLAB platform was used to calculate energy loss spectra by the NS method and is made available through the Mendeley database \citep{REEL_NS_calc}.

\section{Logical flow of the calculations}

\begin{figure}[h!]
    \centering
    \includegraphics[width = 0.98 \textwidth]{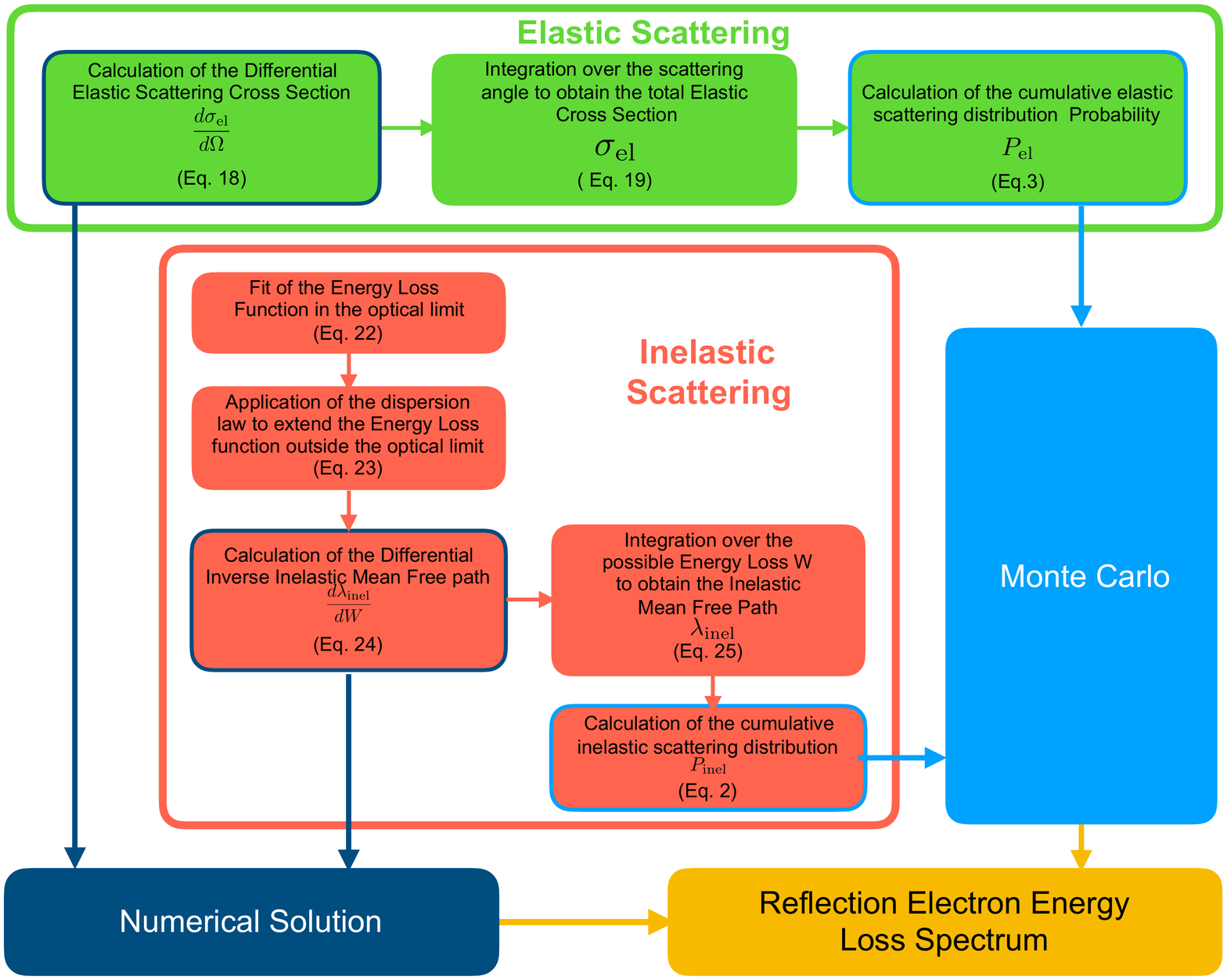}
    \caption{Sketch of the logical flow of REEL spectra simulations.}
    \label{fig:graph}
\end{figure}

To simulate REEL spectra with the MC and NS methods a database, reporting information on elastic and inelastic scattering, has to be inputted to the programs. In Fig. \ref{fig:graph} the computing steps necessary to simulate the energy loss spectra are summarized for both approaches. This scheme refers to equations and calculations that will be presented in the following sections.\\

\textbf{Elastic scattering:} As a first step, the differential elastic scattering cross section $d\sigma_\mathrm{el}/d\Omega$ is calculated for different values of the scattering angle using the following Eq. (\ref{eq:dsigmaEL}). These values represent the input information necessary to perform NS simulations of elastic scattering events. 
To produce the input data used in the MC approach, the differential elastic scattering cross section is integrated over the possible elastic scattering angles to obtain the total elastic scattering cross section $\sigma_\mathrm{el}$ (see Eq. (\ref{elscatcs})). Then, the elastic cumulative distribution probability is evaluated for the full range of possible scattering angles (Eq. (\ref{elcum})). This latter set of data are provided as input information to the MC code suite for calculating the REEL spectrum. 
The whole calculation concerning the elastic scattering is realized by running the MATLAB code {\it Elastic\_calculation.m} in the {\it ESCcal} environment. The code is made publicly available through the Mendeley database \citep{elastic_calc}. 
\\
\par \textbf{Inelastic scattering:}
The description of the inelastic scattering is accomplished by evaluating and fitting the ELF in the optical limit (Eq. (\ref{eq:fit})). The ELF is extended outside the optical limit using dispersion laws as by Eqs. (\ref{fullRPA}, \ref{hqlimit}, \ref{SB}). To obtain the dataset required by the NS approach, the differential inverse inelastic mean free path $d\lambda_\mathrm{inel}/dW$ is calculated for different energy loss ($W$) values (Eq. (\ref{eq:diimfp})).
Moreover, the total inelastic mean free path $\lambda_\mathrm{inel}$ is obtained by integrating over the range of possible energy losses (Eq. (\ref{eq:iimfp})).
Finally, by using Eq. (\ref{inelcum}) the cumulative inelastic scattering distribution is assessed and used as input to the MC routine. 
The inelastic scattering datasets are obtained by running the {\it Inelastic\_calculation.cpp} program, which is provided also via the Mendeley database \citep{inelastic_calc}. 

By providing the relevant information on scattering processes in terms of their probability of occurrence (cross sections) as well as the macroscopic properties of the investigated target (density, atomic mass, atomic number and work function) to the MC and NS code suites, the REEL spectra can be computed. 
A detailed description of the calculation of input data and relative formula is provided in the following sections. 

\section{Calculation of input data}

In the following sections we will provide all the details necessary to calculate the elastic and inelastic cross sections, which are assessed by using the Mott \citep{mott1929scattering,salvat1987analytical, jablonski2004comparison,salvat1993elastic, dapor1996elastic} and the dielectric theory \citep{RitchieHowie,Ritchie_PhysRev_1957}, respectively. 

\subsection{Elastic scattering}\label{ela}

\begin{figure}[h!]
\centering
\includegraphics[width=0.99\textwidth, angle = 0]{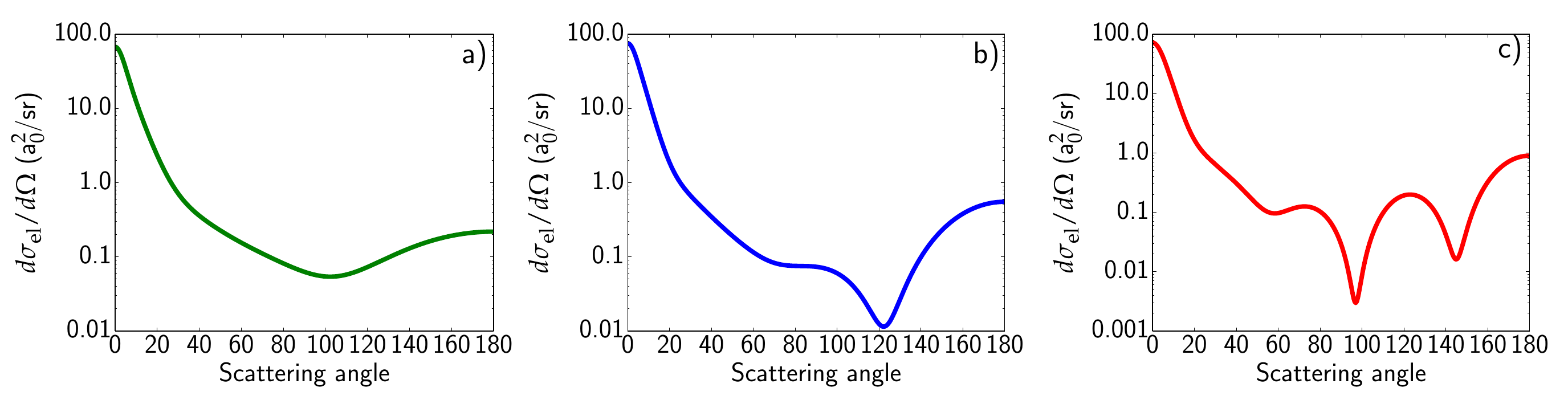}
\caption{DESCSs of a) Cu, b) Ag, and c) Au as a function of the scattering angle. The kinetic energy of the primary beam is set to 1000 eV.\label{fig:DESCS}}
\end{figure}

To simulate the elastic collisions, we solve according to the Mott approach  \citep{mott1929scattering} the Dirac equation in a central field using the partial wave expansion. In particular, for the materials under investigation in this work, the calculation of the elastic scattering cross section was performed using the analytic formulation of the atomic potential proposed by Salvat \cite{Salvat2005,salvat2005elsepa}.
The differential elastic scattering cross section (DESCS) is computed as:
\begin{equation}
    \frac{d\sigma_\mathrm{el}}{d\Omega} = |f(\theta)|^2 + |g(\theta)|^2 ,
    \label{eq:dsigmaEL}
\end{equation}
where $f$ and $g$ are the scattering amplitudes, which describe the asymptotic behaviour of the spherical component of the scattering wave function. 
The DESCSs for copper, silver, and gold calculated by using Eq. (\ref{eq:dsigmaEL}) are reported in Fig. \ref{fig:DESCS} as a function of the scattering angle $\theta$ for a beam kinetic energy of 1000 eV. 

\begin{figure}[h!]
\centering
\includegraphics[width=0.3\linewidth]{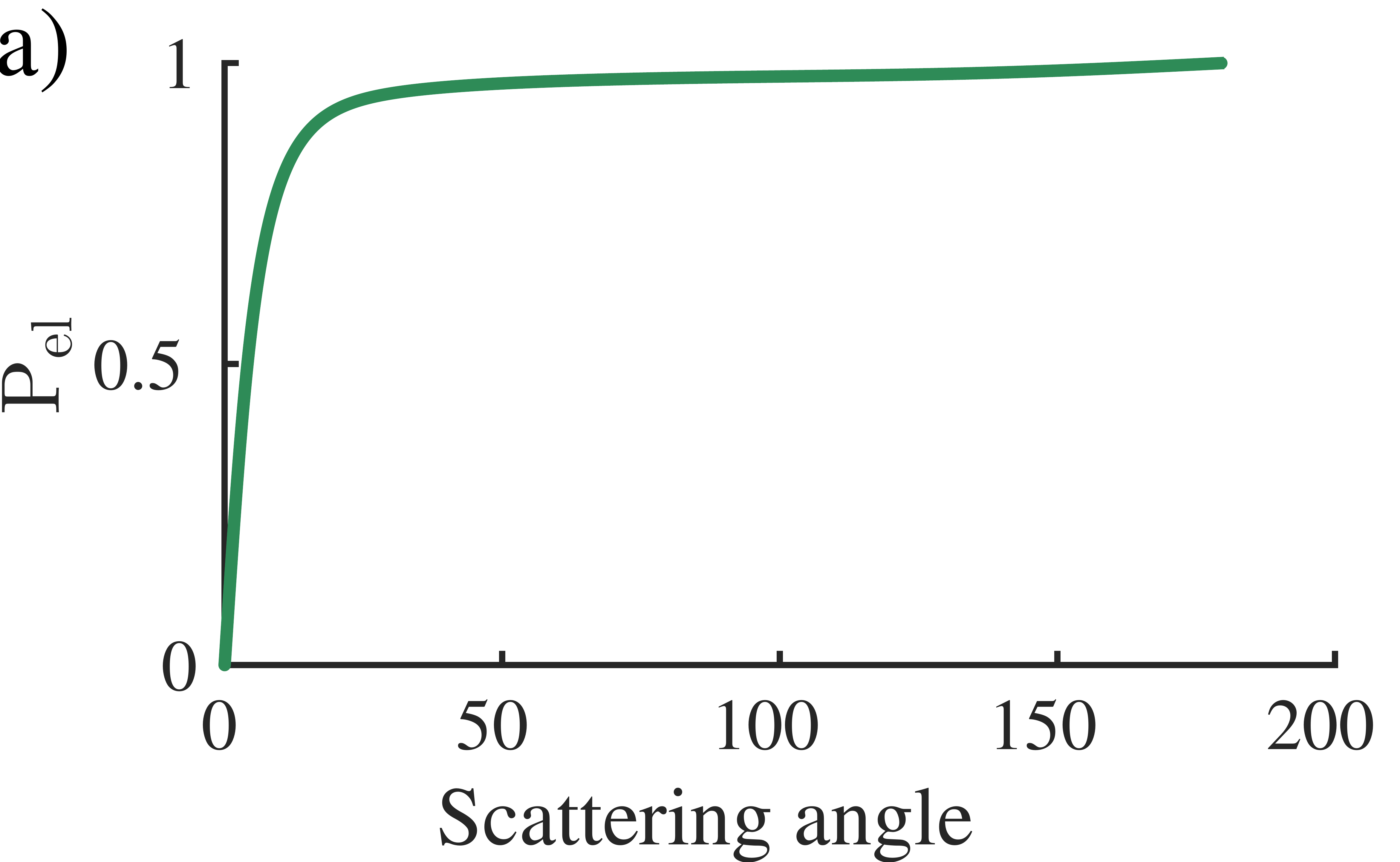}
\hspace{2mm}
\includegraphics[width=0.3\linewidth]{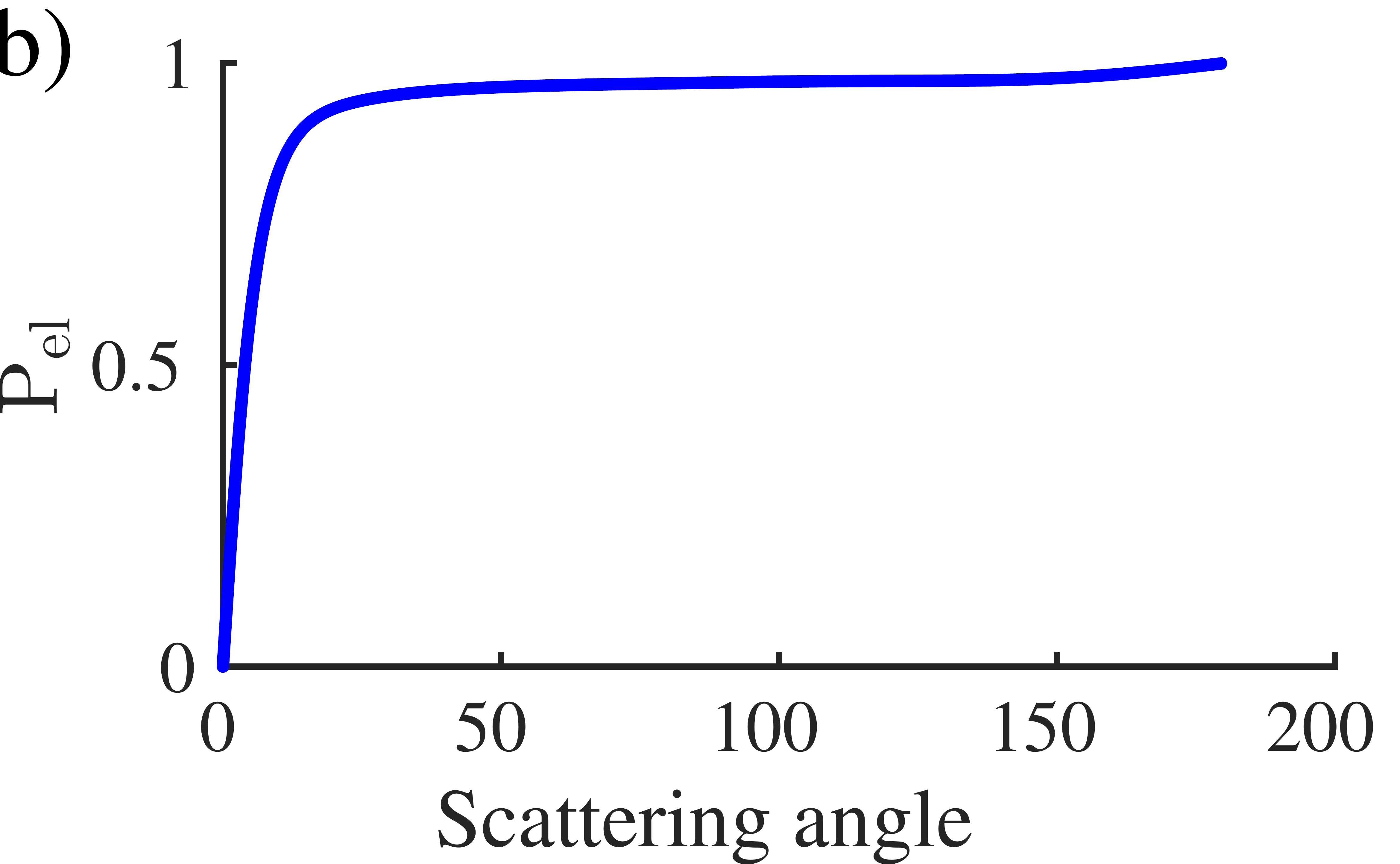}
\vspace{2mm}
\includegraphics[width=0.3\linewidth]{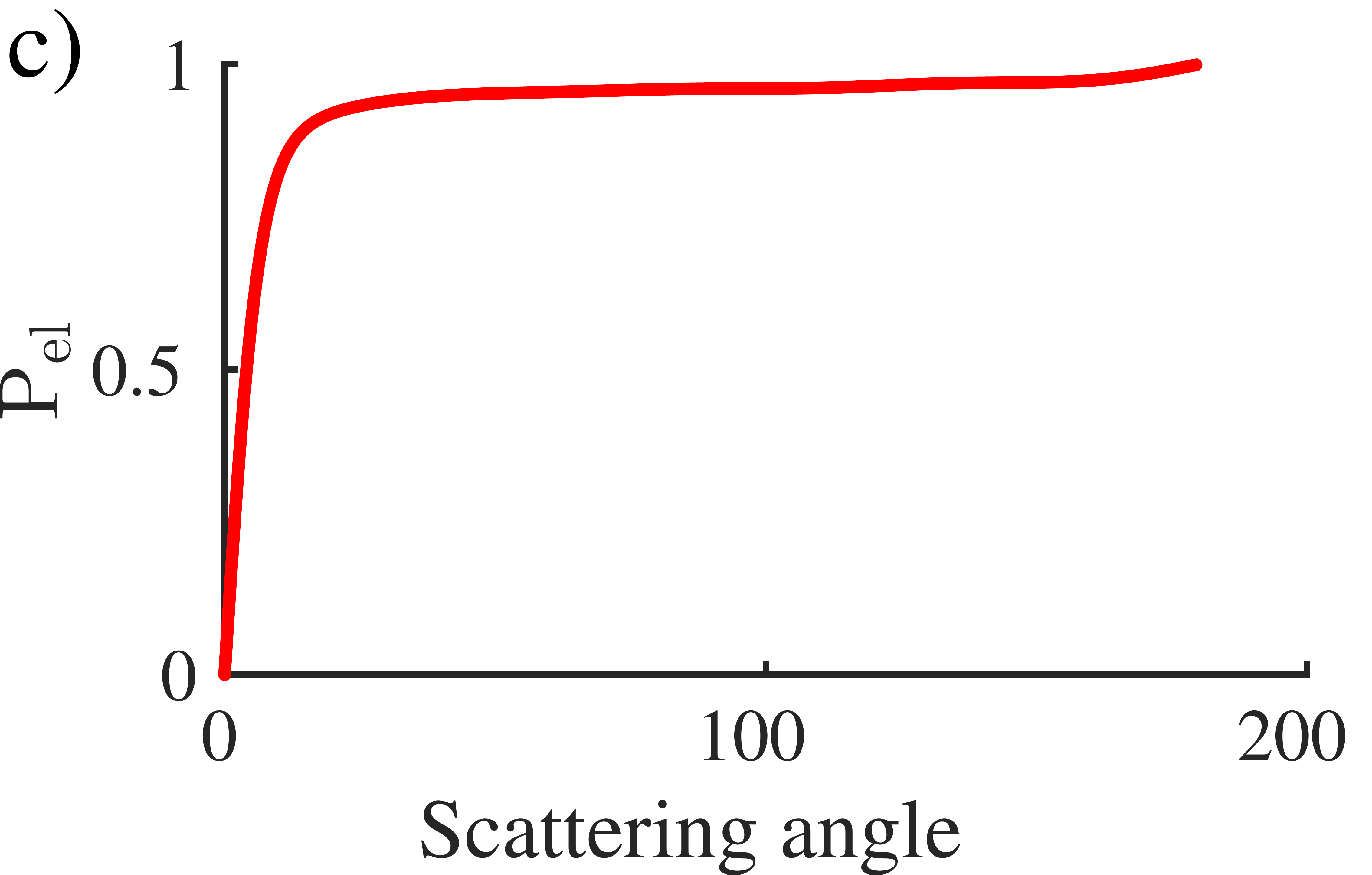}
\caption{Cumulative elastic probability distributions of Cu (a), Ag (b) and Au (c) as a function of the scattering angle. The kinetic energy of the primary beam is set to 1000 eV. \label{fig:pel}}
\end{figure}

The total elastic  scattering cross section (ESCS) is assessed by integrating the DESCS over the solid angle: 
\begin{equation}\label{elscatcs}
\sigma_\mathrm{el} = \int \frac{d\sigma_\mathrm{el}}{d\Omega}d\Omega =2\pi \int\limits_0^\pi \frac{d\sigma_\mathrm{el}}{d\Omega} \sin\vartheta d\vartheta.
\end{equation}
Moreover, the elastic mean free path (EMFP), which is used in both MC and NS simulations, can be computed from Eq. (\ref{elscatcs}):
\begin{equation}
    \lambda_{\mathrm{el}} = \frac{1}{N \sigma_\mathrm{el}},
\end{equation}
where $N$ is the atomic density.\\
\indent Finally, the cumulative elastic probability distribution P$_\mathrm{el}$ is obtained by using Eq. (\ref{elcum}).
The latter are plotted for the three materials under investigation in Fig. \ref{fig:pel} for $E = 1000$ eV.

$\mathrm{P}_\mathrm{el}$ is used in the MC simulations for determining the change in the direction of the beam electrons due to elastic collisions. The trajectory angular deflection is assessed by generating a random number $r$ (uniformely distributed between 0 and 1), and by finding the value of the upper extreme of integration $\overline{\theta}$ in Eq. (\ref{elcum}) that equalizes $r$.

\subsection{Inelastic scattering}\label{inela}

Electron-electron interactions produce energy loss in the primary electron beam. 
This loss can be assessed by  evaluating the dielectric function $\epsilon(\vec{q}, W)$ of the target material as a function of the transferred momentum $\vec{q}$ and energy $W$. 
Within the dielectric theory developed by Ritchie \citep{Ritchie_PhysRev_1957}, the key ingredient is the ELF, which is defined as the negative reciprocal of the imaginary part of the dielectric function: 

\begin{equation}
\mathrm{ELF} = \mathrm{Im}\left[- \frac{1}{\epsilon(\vec{q}, W) }\right].   
\end{equation}
We notice that our system is modeled as a homogeneous and isotropic sample, thus we drop the angular dependence of the momentum transfer of the dielectric function, and we consider only its modulus.\\
\indent In the optical limit ($q\longrightarrow 0$), the ELF can be fitted by a sum of DL oscillators as follows \citep{RitchieHowie}:
\begin{equation}
    \mathrm{Im}\left[ - \frac{1}{\epsilon(q=0, W) }\right] = \sum_n \frac{A_n \Gamma_n W}{(E_n^2 - W^2)^2 + W^2\Gamma_n^2} ,
\label{eq:fit}
\end{equation}
where $A_n$ is the excitation strength of the $n$-th oscillator, $\Gamma_n$ the damping constant, and $E_n$ the excitation energy. \\
\indent The optical ELF is then extended to finite transferred momentum by applying dispersion laws \citep{Kyriakou} according to the presence of bulk or surface plasmon excitation as outlined further below. \\
\indent Moreover, from the ELF, the DIIMFP can be computed as: 
\begin{equation}
    \frac{d \lambda_\mathrm{inel}^{-1}}{dW} = \frac{1}{\pi E a_0} \int\limits_{q_-}^{q+} \frac{dq}{q}\mathrm{Im}\left[-\frac{1}{\epsilon(q, W) }\right],
    \label{eq:diimfp}
\end{equation} 
where  $a_0$ is the Bohr radius and $E$ the electron kinetic energy. The limits of integration in Eq. (\ref{eq:diimfp}) are set to $q_\pm = \sqrt{2mE}\pm \sqrt{2m(E-W)}$ from momentum conservation, while the angular deviation upon inelastic scattering in MC simulations is computed according to the classical binary collision theory, which gives $\sin^2(\theta) = W/E$.
Nevertheless, we notice that the angular pattern of inelastic collisions can also be retrieved by using the momentum-dispersed dielectric function. We remind that the differential inelastic mean free path (DIIMFP) associated with the electron motion within the bulk can be calculated according to the Chen-Kwei theory \cite{tung1994differential,CHEN1996131} equivalently in terms of the momentum transfer or of the polar scattering angle $\theta$. In particular, the inelastic collision angular range can be derived from the asymptotic behaviour of the energy loss dispersion law, which is parabolic for high momentum transfer \cite{doi:10.1002/sia.5495}, and is found to be $[0, \theta_\mathrm{max}]$, where $\theta_\mathrm{max}=\sqrt{\hbar\omega/E}$.
Then, we performed the REEL spectra simulations of Al using our Monte Carlo approach by imposing the two extreme values of the scattering angles to the electrons undergoing inelastic collision $\theta=0$ rad and $\theta=\theta_\mathrm{max}$.
\begin{figure}[h!]
    \centering
    \includegraphics[width = 0.55 \textwidth, angle = 0]{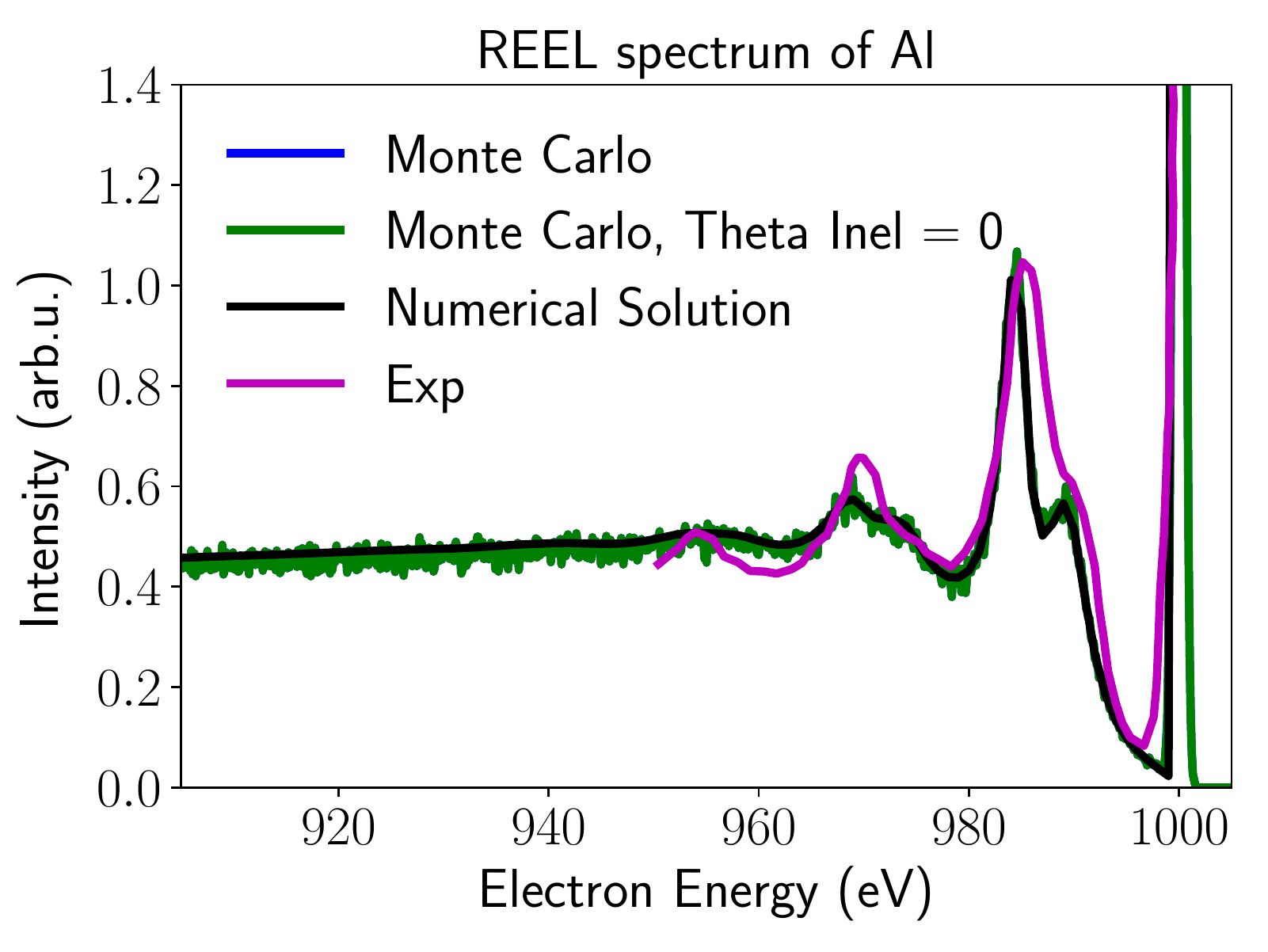}
    \caption{Monte Carlo simulations of the Al REEL spectra for different inelastic scattering angles.}
    \label{fig:Reelangle}
\end{figure}
The results are reported in Fig. \ref{fig:Reelangle} showing that there is no appreciable difference by considering the maximum scattering angle derived by the momentum transfer-energy loss dispersion and the minimum value 0 rad (overlapping black and blue continuous curves in Fig. \ref{fig:Reelangle}). 
On the other hand, the value of the scattering angle that one obtains from the classical binary approach in the small angle approximation must be of course within the range $[0, \theta_\mathrm{max}]$. Thus, modeling the scattering angle by using the binary collision theory can be considered safe with respect to the interpretation of the REEL spectra. This makes the Monte Carlo approach robust with respect to changes in the model used to describe the angular deflection upon inelastic scattering.
\\
\indent Finally, the total IMFP can be obtained by integrating the DIIMFP in the energy loss interval: 
\begin{equation}
    \lambda_\mathrm{inel}^{-1} = \int\limits_{0}^{W_{\mathrm{max}}} \frac{d \lambda_\mathrm{inel}^{-1}}{dW}dW .
    \label{eq:iimfp}
\end{equation}
The upper and lower integration limits are, conventionally, fixed to the energy gap $E_\mathrm{g}$ ($E_\mathrm{g}=0$ in metallic samples) and one half of the initial kinetic energy $E$ plus the energy band gap $E_\mathrm{g}$, respectively. Additionally, the application of Pauli's exclusion principle leads to that energy transferred by an electron of energy $E$ cannot exceed $[E-E_\mathrm{f}]$, where $E_\mathrm{f}$ is the Fermi energy for the metals. These two constraints set an upper bound to the energy transfer in inelastic collision events to $W_\mathrm{max}=\mathrm{min}[(E+E_\mathrm{g})/2, E-E_\mathrm{f}]$.\\
\indent In principle, exchange-correlation effects play an important role in modeling charge transport in solids, as they lead to a variety of inelastic phenomena, such as the polaron-electron interaction. A rigorous treatment within the dielectric formalism of these many-body effects can be achieved in principle by calculating the energy loss function from ab-initio simulations, notably using many-body perturbation theory (GW) or time-dependent density functional theory (TDDFT) \cite{10.1667/RR3281,doi:10.1063/1.4824541}. These methods are available at the expense of high computational cost. While using these ab-initio approaches is out of the scope of the present work, we remind that here the calculation of  the IMFP is based on the dielectric function measured in the optical limit ($q=0$); these experimental data do inherently include exchange-correlation effects. Nevertheless, the inclusion of exchange-correlation effects beyond mean-field is missing at finite momentum transfer, where we extend the optical dielectric function via the DL model. applying, we argue that these effects should have a small impact on the REEL spectra at the energies we deal with in this work ($\approx$ 1 KeV). 
To ground this statement on more quantitative basis we decided to perform the IMFP simulations by including the exchange corrections via the Born--Ochkur model \cite{doi:10.1002/sia.5947} using the full RPA dispersion with both quadratic (see Eq. (\ref{hqlimit})) and linear (see Eq. (\ref{SB})) terms.
\begin{figure}[h!]
\centering
\includegraphics[width=0.3\linewidth, angle = 0]{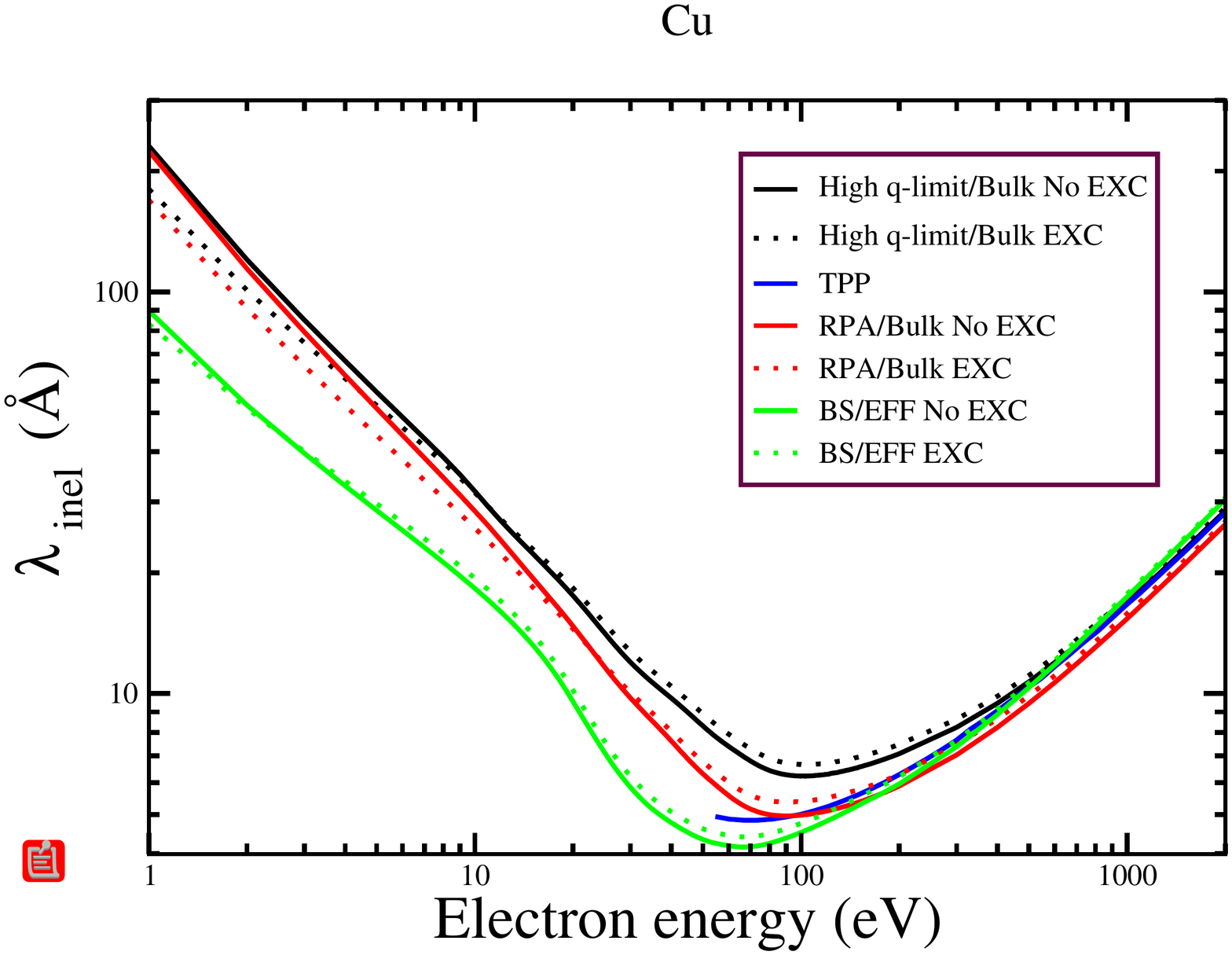}
\includegraphics[width=0.3\linewidth, angle = 0]{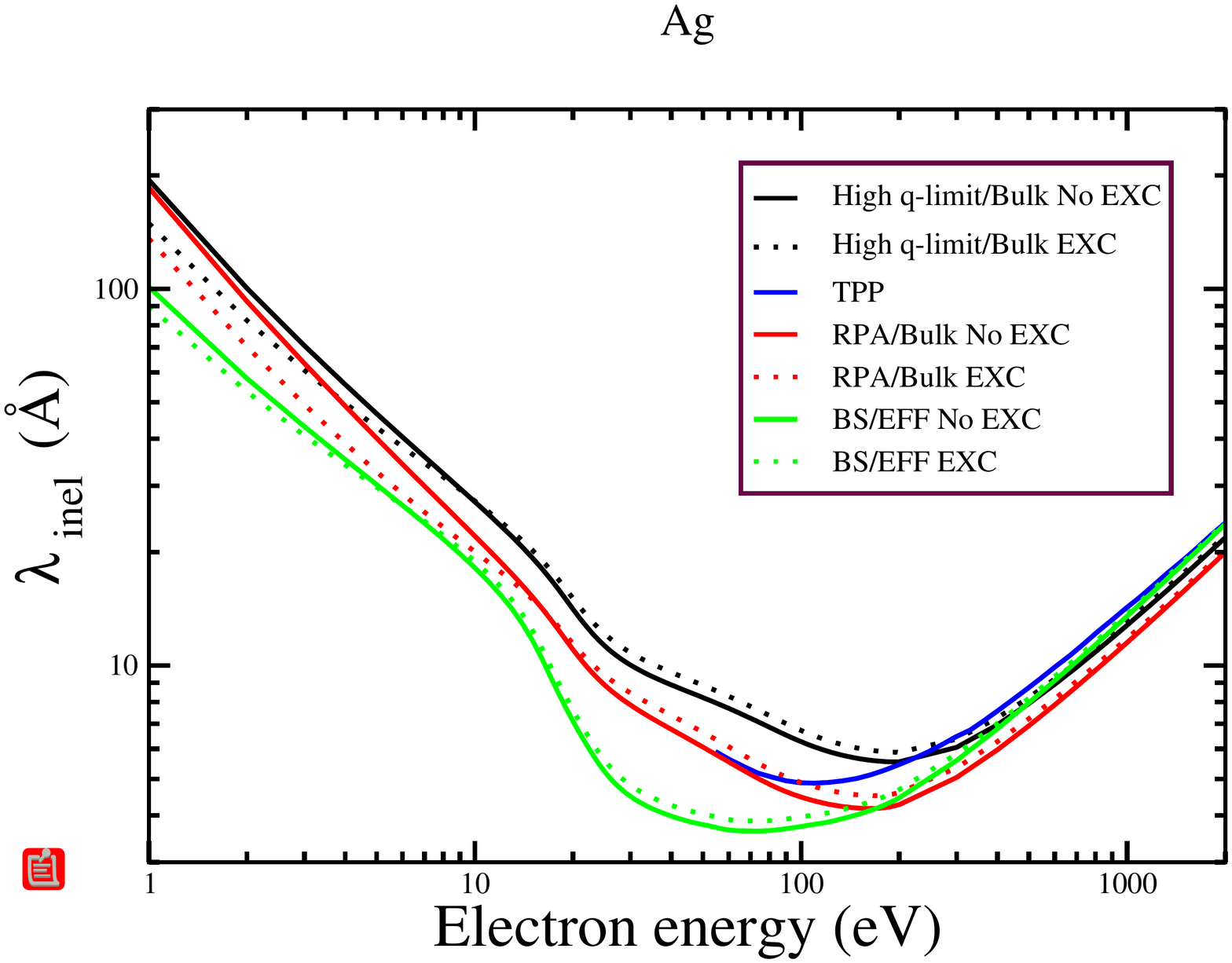}
\includegraphics[width=0.3\linewidth, angle = 0]{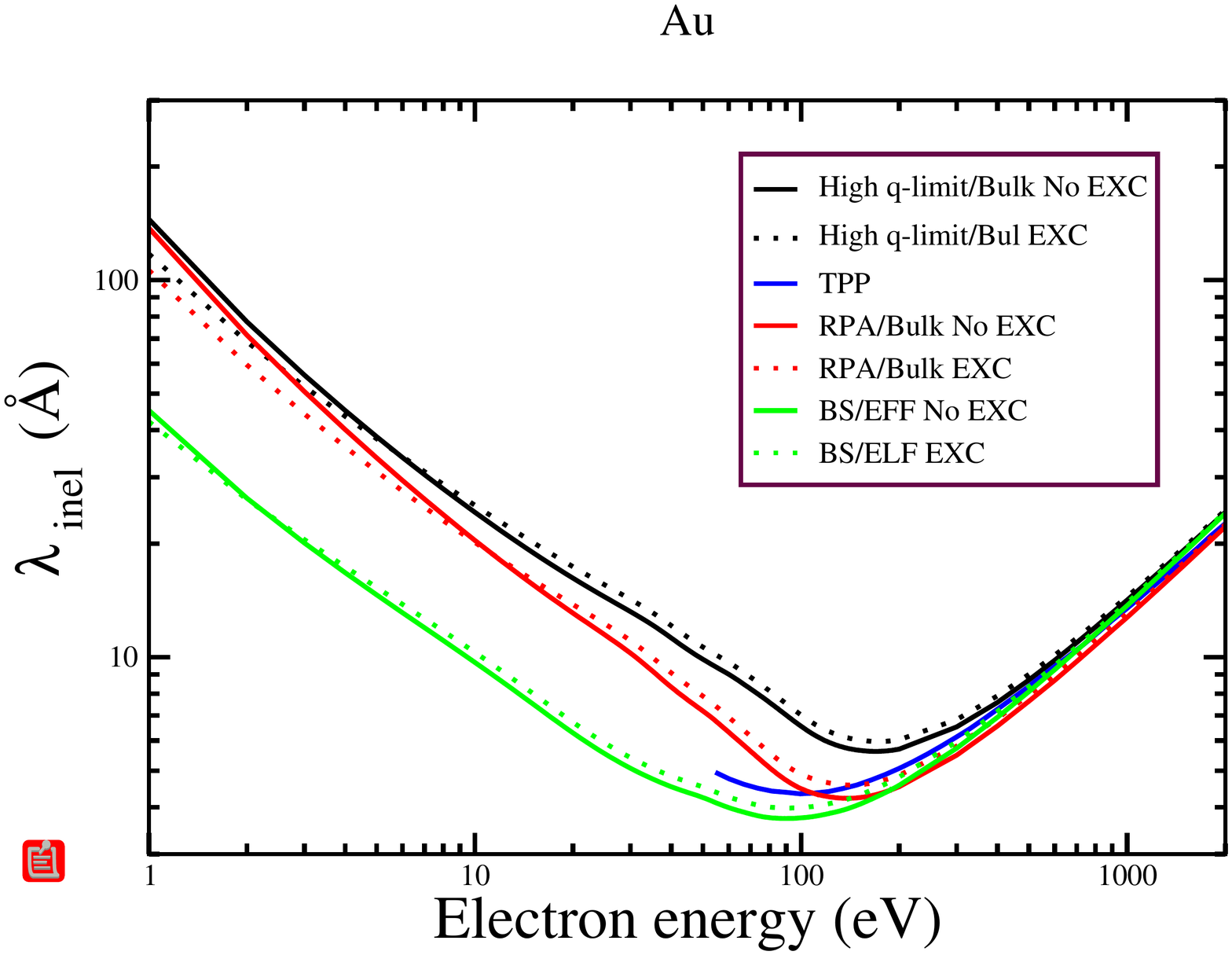}
\caption{IMFPs of a) Cu, b) Ag, and c) Au with (dashed curves) and without (continuous curves) the Born–Ochkur electron exchange corrections in comparison to simulations by Tanuma et al. \cite{tanuma2011calculations} (blue curve). The IMFPs are calculated using respectively the bulk ELFs at high $q$-limit (black curves, see Eq. (\ref{hqlimit})), the full 3D bulk ELF (red curves, RPA, see Eq. (\ref{fullRPA})), and the effective ELF with the bulk-surface extension to finite momenta (green curves, BS, see Eq. (\ref{SB})). Abscissa and ordinate are reported in log-log scale to amplify the difference among the different treatments of the $q$-dispersion. \label{fig:lambda_inel}}
\end{figure}
Within this approximation, exchange effects are modelled as follows:
\begin{equation}
f_\mathrm{exc}(k)=\left(\frac{\hbar q}{m v}\right)^4- \left(\frac{\hbar q}{m v}\right)^2\,,
\end{equation}
where $m$ and $v$ are the electron mass and velocity, respectively.
In Fig. \ref{fig:lambda_inel} we plot the IMFPs for Cu, Ag, and Au by (dashed curves) and by not (continuous curves) including the exchange and correlation corrections for different dispersion laws to finite momenta (see sections 4.2.1 and 4.2.2). We notice that the differences are not significant, particularly when the energies are higher than $\sim$ 100 eV, considering that the figures are plotted in log-log scale. Thus, the REEL spectra have been eventually simulated without including this exchange correction.\\
\indent Furthermore, we notice that in a previous work \cite{azzolini2017monte} we carried out the calculation of the frequency and momentum dependent dielectric response of diamond and graphite in two ways: a full ab initio approach, in which we carry out time-dependent density functional simulations in linear response for different momentum transfer vectors, and a semi-empirical extended DL model. Ab initio calculated dielectric functions of these two carbon-based materials lead to a better agreement with experimental data of REEL spectra, inelastic mean free path, and stopping power, more significantly in the low energy regime ($<$ 100 eV) with respect to the widely used DL model. \\
\indent Nevertheless, while these discrepancies are particularly evident for insulators and semiconductors beyond the optical limit ($q \neq 0$), where single particle excitations and excitonic effects become relevant, less dramatic effect on the accuracy of MC simulations was found for semi-metals, such a graphite. Indeed, metallic system are more effective in screening excitons and charges than semiconductors. Here we deal with all metallic systems, thus it is not surprising that exchange effects do not play a paramount role in the interpretation of the REEL spectra, particularly  at high energy ($\geq$ 1 keV), where the inelastic features are independent of these effects to all extents and purposes.
\\
\indent We applied this scheme for interpreting the REEL spectra of Cu, Ag and Au. First, we evaluate the ELFs describing only bulk plasmon excitations. Second, the effective ELFs to account for both surface and bulk plasmons are presented. 

\subsubsection{Bulk ELF}

The bulk ELFs  in the optical limit were obtained by using the best-fit parameters provided by Denton {\it et al.}~\citep{denton2008influence} and C.C. Montanari {\it et al.} \citep{montanari2007calculation} for Au and Cu, while in the case of Ag the ELF was obtained by best fitting the optical data from Smith {\it et al.} \citep{smith1985handbook}.
\begin{figure}[h!]
\centering
\includegraphics[width=0.99\textwidth, angle = 0]{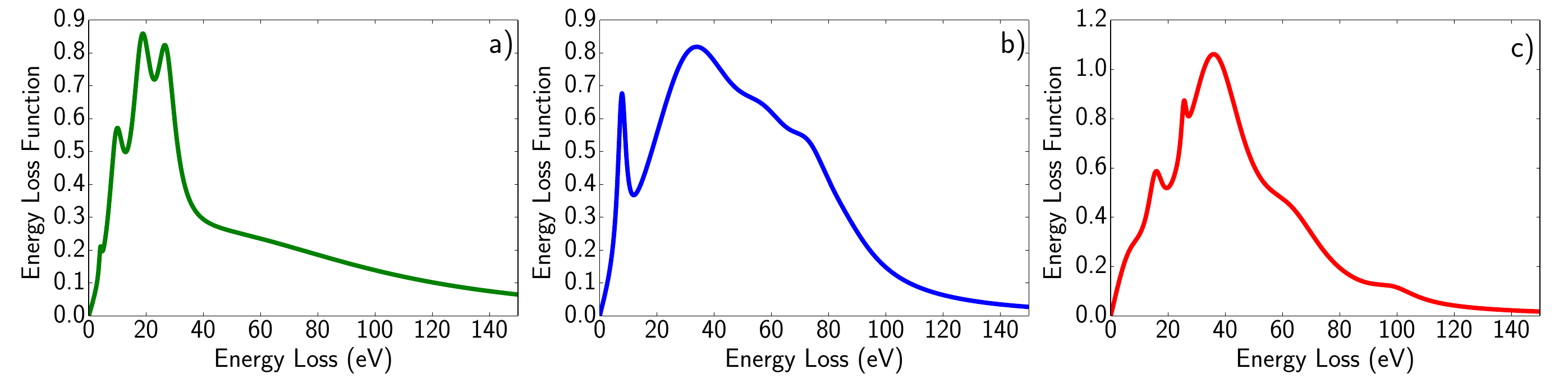}
\caption{Bulk energy loss functions of a) Cu, b) Ag, and c) Au in the optical limit ($q\longrightarrow 0$) obtained using Eq. (\ref{eq:fit}). These ELFs were obtained by best-fitting the optical data reported in \citep{smith1985handbook}  for Ag, while we used the parameters from \citep{montanari2007calculation} and \citep{denton2008influence}  for Cu and Au, respectively. \label{fig:ELF}}
\end{figure}
ELFs of these metals are represented in Figs. \ref{fig:ELF}a), b), and c).

\begin{figure}[h!]
\centering
\includegraphics[width=0.9\textwidth]{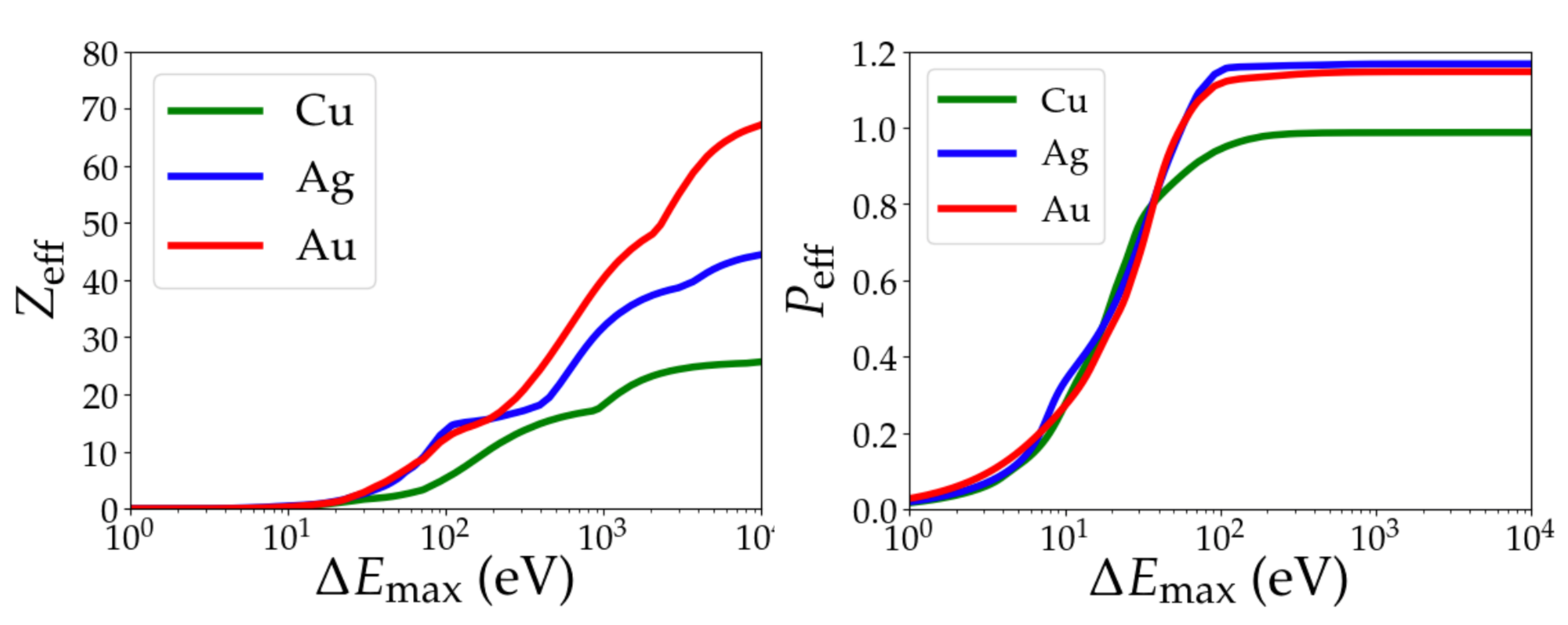}
\caption{Plot of $Z_{\mathrm{eff}}$ (left panel) and $P_{\mathrm{eff}}$ (right panel) versus $\Delta E_{\mathrm{eff}}$ (transferred energy) for Al, Cu, and Ag.\label{fig:PZeff}}
\end{figure}

\begin{table}[h!]
    \centering
    \small\addtolength{\tabcolsep}{-2pt}
    \begin{tabular}{ |c c c c|c c c c|c c c c|}
        \hline
         && Cu &&  & & Ag & & && Au &\\
        \textit{n} & $E_n$ (eV) & $\Gamma_n$ (eV) & $A_n$ (eV$^2$) & \textit{n} & $E_n$ (eV) & $\Gamma_n$ (eV) & $A_n$ (eV$^2$) & \textit{n} & $E_n$ (eV) & $\Gamma_n$ (eV) & $A_n$ (eV$^2$) \\ \hline 
             1 &  4.08	& 1.09	& 0.33 & 1 & 7.89 &	3.37 & 	12.80 & 1 & 9.52	& 14.97	& 18.49\\
             2 & 10.07	& 5.99	& 22.10 & 2 & 38.20 &	42.93 &	1109.46 & 2 & 15.92	& 6.26	& 25.85\\
             3 & 19.05	& 8.16	& 88.91  & 3 & 59.58	& 29.93	& 480.38 & 3 & 25.58	& 2.18	& 11.12\\
             4 & 27.21	& 8.16	& 112.54 & 4 & 73.81	& 20.12	& 300.6& 4 & 38.09	& 26.67	& 973.52 \\
             5 & 78.91	& 152.38 & 2216.74 & 5 & 85.70	& 27.70	&226.83  & 5 & 64.49	& 30.48	& 507.39\\
            & & & & & & & & 6 & 99.32	& 19.05	& 88.88 \\
            & & & & & & & & 7 & 402.71 & 612.23 & 337.32\\
            \hline
    \end{tabular}
     \caption{Fit parameters of Eq. (\ref{eq:fit}) obtained for bulk ELFs. The best-fit parameters of Cu were provided by C.C. Montanari {\it et al.} \citep{montanari2007calculation}, while for
    Au were given by C.D. Denton {\it et al.} \citep{denton2008influence}. In the case of Ag the parameters were obtained by best-fitting the optical measurements by Smith et al \citep{smith1985handbook}. \label{tab:fit}}
\end{table}
The fit parameters are reported in table \ref{tab:fit}.
\\
\indent We performed a check to test that our ELFs satisfy the $f$-sum rule, for which the integral of the ELF multiplied by the energy loss must sum up to an effective number of electrons per atom.
Moreover, we have tested the accuracy of our bulk optical data by using also the perfect-screening
sum-rule (ps--sum or $P_{\mathrm{eff}}$), which should reach unity for energy going to infinity. To calculate the integrals
above $\Delta E_{\mathrm{max}} \ge 72.4$ eV we used the optical data reported in Ref. \citep{HENKE1993181}. We notice that
our results are in good agreement with those presented by Tanuma et al. \citep{doi:10.1002/sia.740111107} and within the error bars there reported.
In Fig. \ref{fig:PZeff} we plot the $Z_{\mathrm{eff}}$ (left panel) for the ELFs of Cu, Ag, and Au, while in the right panel we report the $P_{\mathrm{eff}}$. 
The $Z_{\mathrm{eff}}$, $P_{\mathrm{eff}}$ limiting values are also reported in table \ref{tab:val}.
\begin{table}[h!]
    \centering
    \small\addtolength{\tabcolsep}{-2pt}
    \begin{tabular}{ |c|c c|c c|c c|}
        \hline
         &$~~~~~$ Cu & & ~~~~ Ag  & & ~~~ Au & \\
         $E$ (eV) &$Z_{\mathrm{eff}}$ &  $P_{\mathrm{eff}}$ & $Z_{\mathrm{eff}}$ &  $P_{\mathrm{eff}}$ & $Z_{\mathrm{eff}}$ &  $P_{\mathrm{eff}}$ \\ \hline 
         1 &    5.52e-05 & 0.017 & 9.24e-05 & 0.020 & 1.29e-04 & 0.028\\
          10&     0.09 & 0.275 & 0.17 & 0.336 & 0.1 & 0.27\\
          108.5 & 7.28 & 0.950 & 18.63  & 1.056 & 16.65 & 1.021\\
          1012   &   19.68 & 0.987 & 36.36 & 1.067 & 44.17 & 1.046\\
          9886.4 &   27.07 & 0.987 & 45.46 & 1.067 & 70.74 & 1.047\\
          29779 &   28.23 & 0.987 & 46.51 & 1.067 & 78.94 & 1.047\\
            \hline
    \end{tabular}
   \caption{$Z_{\mathrm{eff}}$, $P_{\mathrm{eff}}$ limiting values of Cu, Ag, and Au represented in Fig. (\ref{fig:PZeff}). \label{tab:val}}
\end{table}

The use of the optical ELF from experimental measurements is justified simply on empirical grounds, by comparing ``a posteriori'' our simulations with the experimental data. 
Nevertheless, to ground this observation on more firm quantitative basis we demonstrate that the discrepancies obtained by using different dispersion laws are not significant with respect to the assessment of the IMFPs in the energy range considered in this work (
$\approx$ 1000 eV modulo the plasmon energy). Indeed, the kinetic energy of the primary beam lies well above the valence excitation of the metallic systems (a few eV). 
The most general dispersion law derived within the RPA of the 3D electron gas to extend the experimental optical data ($q=0$) to finite momentum transfer for bulk plasmons is the following \cite{Kyriakou}
\begin{equation}\label{fullRPA}
\hbar^2\omega^2=  \hbar^2\omega_{np}^2+\beta\frac{\hbar^2q^2}{m}+  \frac{\hbar^4q^4}{4m^2}\,,
\end{equation}
where $\beta=6E_\mathrm{f}/5$ ($E_\mathrm{f}$ is the Fermi energy of the material), $\omega_{np}$ is the plasmon $n$-peak energy, and $m$ is the electron mass. We notice that this expression asymptotically approaches the parabolic single-particle dispersion, as well as the limit value $\omega(q\approx 0)\approx \omega_{np}$ for negligible momentum transfer. At high momentum transfer, in particular, one obtains 
\begin{equation}\label{hqlimit}
\hbar\omega=  \hbar\omega_{np}+\frac{\hbar^2q^2}{2m}\,.
\end{equation}
Moreover, for intermediate $q$, one cannot neglect the quadratic term in the dispersion law for bulk plasmons Eq. (\ref{fullRPA}). \\
\indent To determine quantitatively the impact of different dispersion laws on the IMFP, in Fig. \ref{fig:lambda_inel} we report this quantity for all metals under investigation using on the one hand the full (RPA, see Eq. (\ref{fullRPA})) and on the other hand the asymptotic (high $q$-limit, see Eq. (\ref{hqlimit})) dispersion models of the bulk ELF with (dashed line) and without (continuous line) exchange corrections. We notice again that the figures are in log-log scale to enhance the differences between the two dispersion models. These findings demonstrate that the impact on the IMFP of different dispersion laws for infinite 3D media, particularly in our energy range of 1 keV, is not significant.
This actually was also found previously (see Ref.  \cite{doi:10.1002/sia.5495}). In that case a comparison between Monte Carlo REEL spectra calculated taking either the quadratic or quartic dispersion laws did not find relevant differences in both relative intensities and peak energy positions of both bulk and surface plasmons between these two limiting cases.
\\
\indent Additionally, by using the dispersion law for the transferred momentum of Eq. (\ref{fullRPA}), DIIMFPs are calculated as by Eq. (\ref{eq:diimfp}), which can be integrated to obtain the IMFPs of Cu, Ag, and Au.

\subsubsection{Effective ELF for including surface plasmons}

\begin{figure}[h!]
\centering
\includegraphics[width=0.96\textwidth, angle = 0]{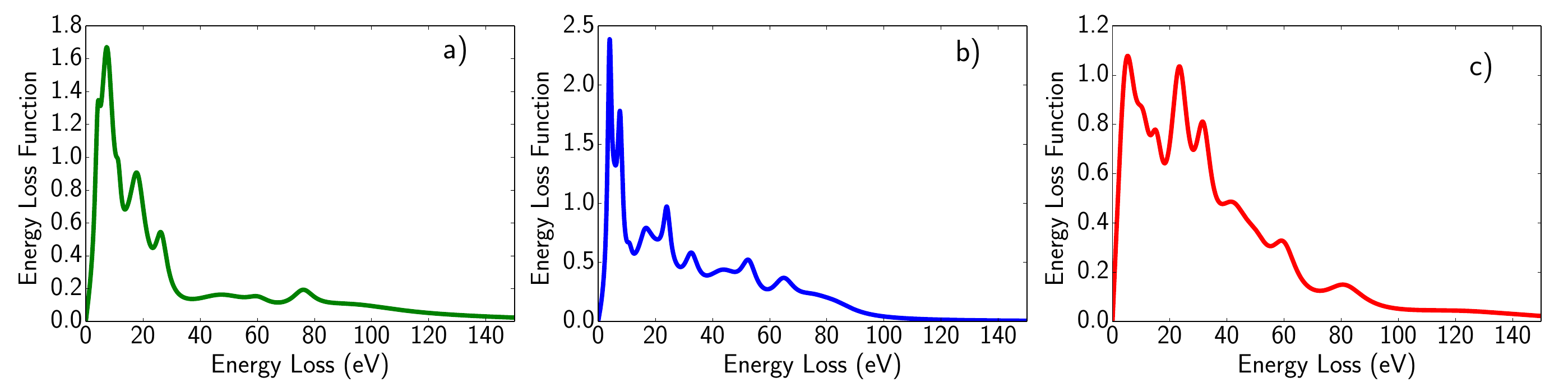}
\caption{Effective energy loss functions of Cu a), Ag b), and Au c) in the optical limit ($\vec{q}
\longrightarrow 0$) obtained using Eq. (\ref{eq:fit}). These ELFs were obtained by best-fitting the effective ELFs calculated by Nagatomi et al \citep{nagatomi2003construction}. \label{fig:ELFeff}}
\end{figure}

\begin{figure}[h!]
\centering
\includegraphics[width=0.99\textwidth, angle = 0]{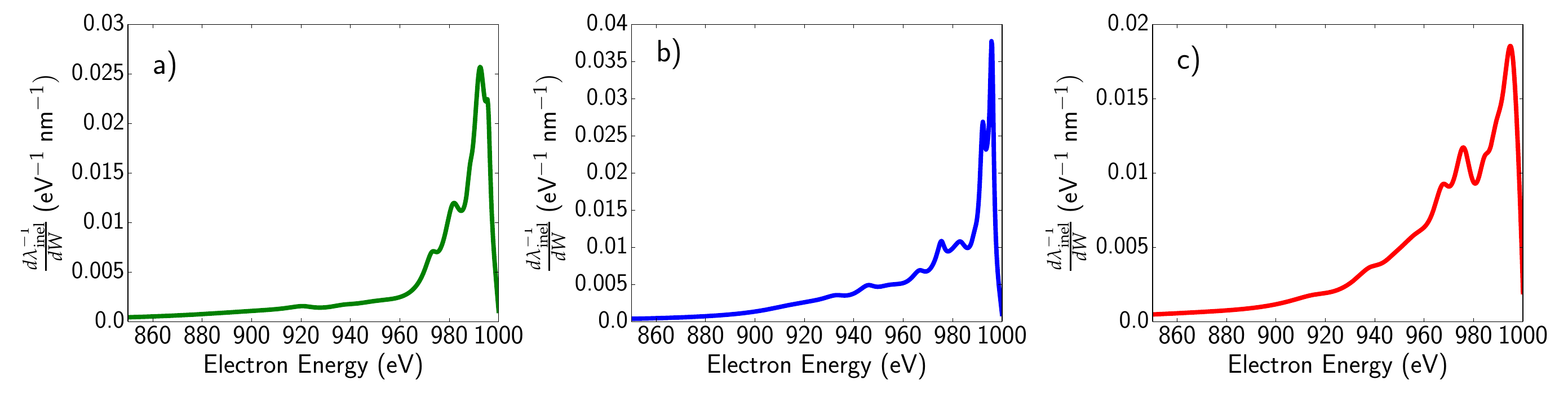}
\caption{DIIMFPs of Cu a), Ag b) and Au c) calculated using Eq. (\ref{eq:diimfp}) and the effective ELFs, for a kinetic energy of the primary beam equal to 1000 eV. \label{fig:diimfEFF}}
\end{figure}

An extension of the previous approach to assess bulk ELFs is represented by the inclusion of surface plasmons in the description of the dielectric properties and, thus, in the interpretation of REEL spectra. 
In general, the inclusion of surface scattering is crucial to obtain an accurate evaluation of the number of electrons emerging out of the solid and thus it is of paramount importance in the study of secondary electron generation \cite{azzolini2018secondary}. The effective ELF allows us to treat equally bulk and surface inelastic scattering events leading to energy loss, including effectively the surface plasmons in the description of REEL spectra. Compared to the bulk ELFs derived by electron transmission measurements, this approach lumps together the information on bulk and surface (and interface) excitation within an effective ELF. \\
\indent The effective ELFs of Cu, Ag and Au were obtained by Nagatomi {\it et al.} \citep{nagatomi2003construction} from experimental REEL spectra by applying the extended Landau theory \citep{yoshikawa1995h}.
In particular, we have fitted this effective optical ELF by using Eq. (\ref{eq:fit}). This generalizes the DL procedure used previously to include surface plasmons. The best-fit parameters and the ELFs are reported respectively in table \ref{tab:fitEFF} and Fig. \ref{fig:ELFeff}. \\ 
\indent Additionally, the use of the effective ELF to include surface collective excitation leads to a dispersion law different from the bulk (see Eqs. \ref{fullRPA}, \ref{hqlimit}), which reads \cite{Kyriakou}:
\begin{equation}\label{SB}
\hbar^2\omega^2=  \hbar^2\omega_{np}^2+\hbar^2\alpha q+\beta\frac{\hbar^2q^2}{m}+  \frac{\hbar^4q^4}{4m^2}\,,
\end{equation}
where
$\alpha=\sqrt{3E_f/(5m)}\omega_p$ and $\omega_p=\sqrt{4\pi ne^2/m}$ is the bulk plasmon nominal energy ($e$ is the electron charge). \\
\indent The presence of a quasi 2D slab on top of the semi-infinite bulk results in the addition of a linear term in the momentum transfer dispersion law Eq. (\ref{SB}). We also performed the calculation of the IMFP for all metals under investigation by using this bulk-planar surface model (BS) and the results are plotted in Fig. \ref{fig:lambda_inel}. Again the seemingly different behaviour is due to the use of the log-log scale in the plot, which enhances intentionally the discrepancies, otherwise negligible above 1 keV.
\begin{table}[h!]
    \centering
    \small\addtolength{\tabcolsep}{-2pt}
        \begin{tabular}{ |c c c c|c c c c|c c c c|}
        \hline
         Cu && && Ag & && &Au &  & &\\
        \textit{n} & $E_n$ (eV) & $\Gamma_n$ (eV) & $A_n$ (eV$^2$) & \textit{n} & $E_n$ (eV) & $\Gamma_n$ (eV) & $A_n$ (eV$^2$) & \textit{n} & $E_n$ (eV) & $\Gamma_n$ (eV) & $A_n$ (eV$^2$) \\ \hline 
        1 & 	 4.28 & 2.09 &  5.80  & 1  & 4.09 &  1.70 &	 13.80 & 1 & 6.40 & 8.77 & 	 46.92\\ 	 
        2 & 7.90& 	 5.79& 	 67.00 &2 & 5.91 &  	 2.80 &  8.42 & 2 & 11.11 &	 6.52 &  	26.16 \\
        3 & 11.45&  	 2.20& 	 6.50 & 3& 7.73 &	 2.13 &	21.16& 3 &  15.53 &	 5.62 & 30.56  \\ 
        4 & 14.76& 	 5.20 & 	  5.20& 4& 10.98 &	 2.30 & 	3.72 &4 & 23.68 &  	 7.72 & 148.97\\
        5 & 18.30&	 7.16& 95.00 &5 &16.58 &	 4.80 &	15.80 &5 &31.90 &	 6.90&	110.32\\
        6 & 26.51 & 	 5.16& 	 50.12& 6& 20.58& 	 18.20& 	190.83 &6& 42.82 & 	 15.60 & 220.60	\\ 
        7 & 48.91 &	 25.30& 	 146.00 &7&24.10 	& 2.90 &	33.60 &7& 50.51 &	 11.20& 65.50\\
        8 & 60.30 & 	 9.30& 	 28.20&8&32.75 & 	 5.60&    53.60	&8& 60.12& 	 10.70& 	132.50 \\
        9 &76.20 &	 9.00 &	 75.20&9&44.48&	 16.20&	220.30 &9&81.65 &	 16.78 &	143.20 \\ 
        10& 95.20& 45.00 & 	330.00&10&52.75 &	 6.50& 	  102.20&10&125.60 &	 60.30 &	250.50 \\
        11 & 130.40 &	 90.30 	&  180.20 &11& 65.09 & 9.20 & 138.50 & & & & \\
        & & & & 12& 75.21&  16.50 & 140.70 & & & & \\
        & & & & 13& 83.15 &	17.30 & 120.50 & & & & \\ \hline

        \end{tabular}
        \caption{The best-fit parameters (Eq. (\ref{eq:fit})) of effective ELFs of Cu, Ag and Au. \label{tab:fitEFF}}
\end{table}
Furthermore, by applying the dispersion law for finite transferred momenta (Eq. (\ref{SB})), we assessed the DIIMFP as by Eq. (\ref{eq:diimfp}) using the effective ELFs. The DIIMFPs for Cu, Ag and Au are reported respectively in Fig. \ref{fig:diimfEFF}a), b) and c) for an initial kinetic energy of the beam equal to 1000 eV. 
By integrating the DIIMFs over the possible energy losses (Eq. (\ref{eq:iimfp})) the IIMFP is obtained (see Fig. \ref{fig:lambda_inel}).\\
\indent Moreover, the cumulative inelastic probabilities were calculated as reported in Eq. (\ref{inelcum}). P$_\mathrm{inel}(E,W)$, along with the IMFP, are used as input data in the MC simulations of the REEL spectra for the three materials under investigation.\\
\indent Finally, to determine quantitatively the impact of the different dispersion laws that characterize the 3D bulk and the bulk-planar surface, we also report in the following sections the REEL spectra using both the bulk and the effective ELFs in MC and NS approaches, respectively, along with their relevant dispersion laws.
Furthermore, we discuss more extensively the impact of bulk and surface  contributions on REEL spectra by reporting simulations of Al, where the bulk and surface plasmon peaks are well resolved in energy.\\
\indent We stress that also the NS method requires the knowledge of the DIIMFP and of the IMFP to compute the REEL spectra.

%%%%%%%%%%%%%%%%%%%%%%%%%%%%%%%%%%%%%%%%%%%%%%%%%
\section{Results and discussion}

MC and NS calculations were performed by fixing the primary electron beam kinetic energy $E_0=$ 1000 eV for all the different targets (Cu, Ag and Au). In our simulations we assumed normal incidence of the primary beam and collected all the electrons emitted from the surface. Thus, our spectra are integrated between 0$^\circ$ to 90$^\circ$ with respect to the normal to the surface and all over the azimuth angle.
In MC simulations we used 10$^9$ trajectories, in order to have a reliable statistics for calculating the REEL spectra.
While our MC approach is of course able to track the electrons within the solid down to their ultimate cut off energy (the electron affinity $\chi$, see Ref. \cite{azzolini2018secondary}), here we do not calculate secondary electron spectra and the simulated REEL spectra energy range ($\approx$ 1 keV) is well above this cut off. Thus, the energy cut-off to simulate the REEL spectra was set to 900 eV, down from the elastic peak at 1 keV.\\
\indent In the NS approach $K$ is the number of inelastic scattering events that the electrons undergo before being considered at rest. A value of $\bar{K}=25$ is large enough to ensure this. In this respect in Fig. \ref{fig:reelNin} we report the REEL spectra deconvoluted for the number of inelastic collisions that electrons undergo in their way out of the solid. One can see that for a number of inelastic collisions $\bar{K}=5$ the REEL spectra of Cu, Ag and Au are already indistinguishable from those having inelastically scattered more than 25 times ($\bar{K}=25$). Thus the results must be considered safely at convergence for that large value.\\
\indent The samples characteristic parameters used in the calculations are summarized in table \ref{tab:el}. 

\begin{table}[h!]
    \centering
    \small\addtolength{\tabcolsep}{-3pt}
    \begin{tabular}{|c|c c c|}
    \hline
    &  Cu& Ag & Au \\\hline
    density (g/cm$^3$) & 8.96 \citep{montanari2007calculation}& 10.5\citep{tanuma2011calculations} & 19.32\citep{denton2008influence}  \\
    $\lambda_\mathrm{el}$ (nm) & 0.866& 0.783 & 0.677 \\
    $\lambda_\mathrm{inel}$ (Bulk ELF) (nm)  & 1.686 & 1.277 & 1.414\\
    $\lambda_\mathrm{inel}$ (Effective ELF) (nm)  &1.713& 1.369 & 1.418\\
    \hline
    \end{tabular}
    \caption{Target material characteristic parameters. Elastic and inelastic mean free paths have been assessed at $E_0=$ 1000 eV primary beam kinetic energy. }
    \label{tab:el}
\end{table}

\subsection{Bulk ELF: REEL spectra}

\begin{figure}[h!]
\centering
\includegraphics[width=0.3\linewidth, angle = 0]{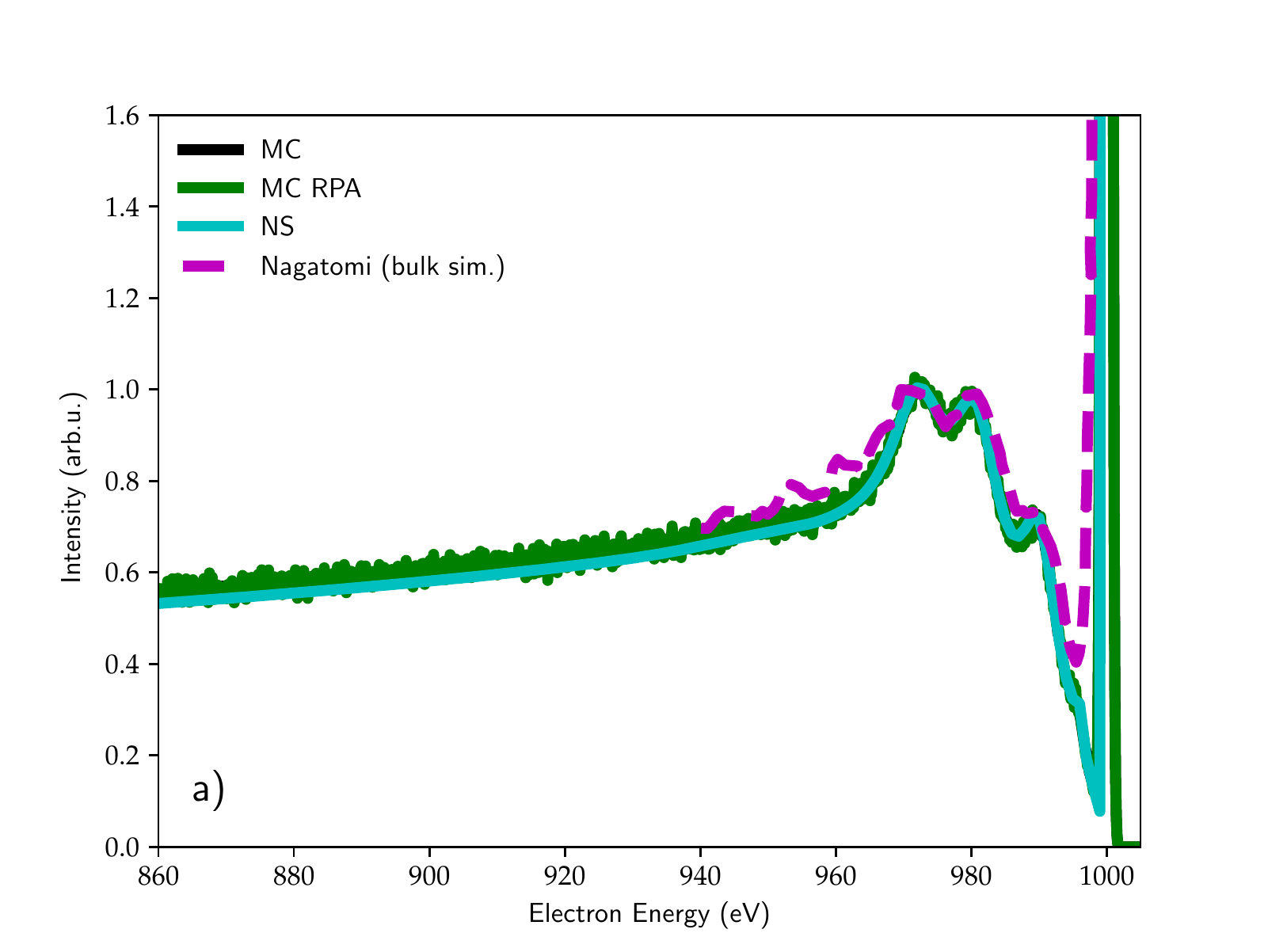}
\includegraphics[width=0.3\linewidth, angle = 0]{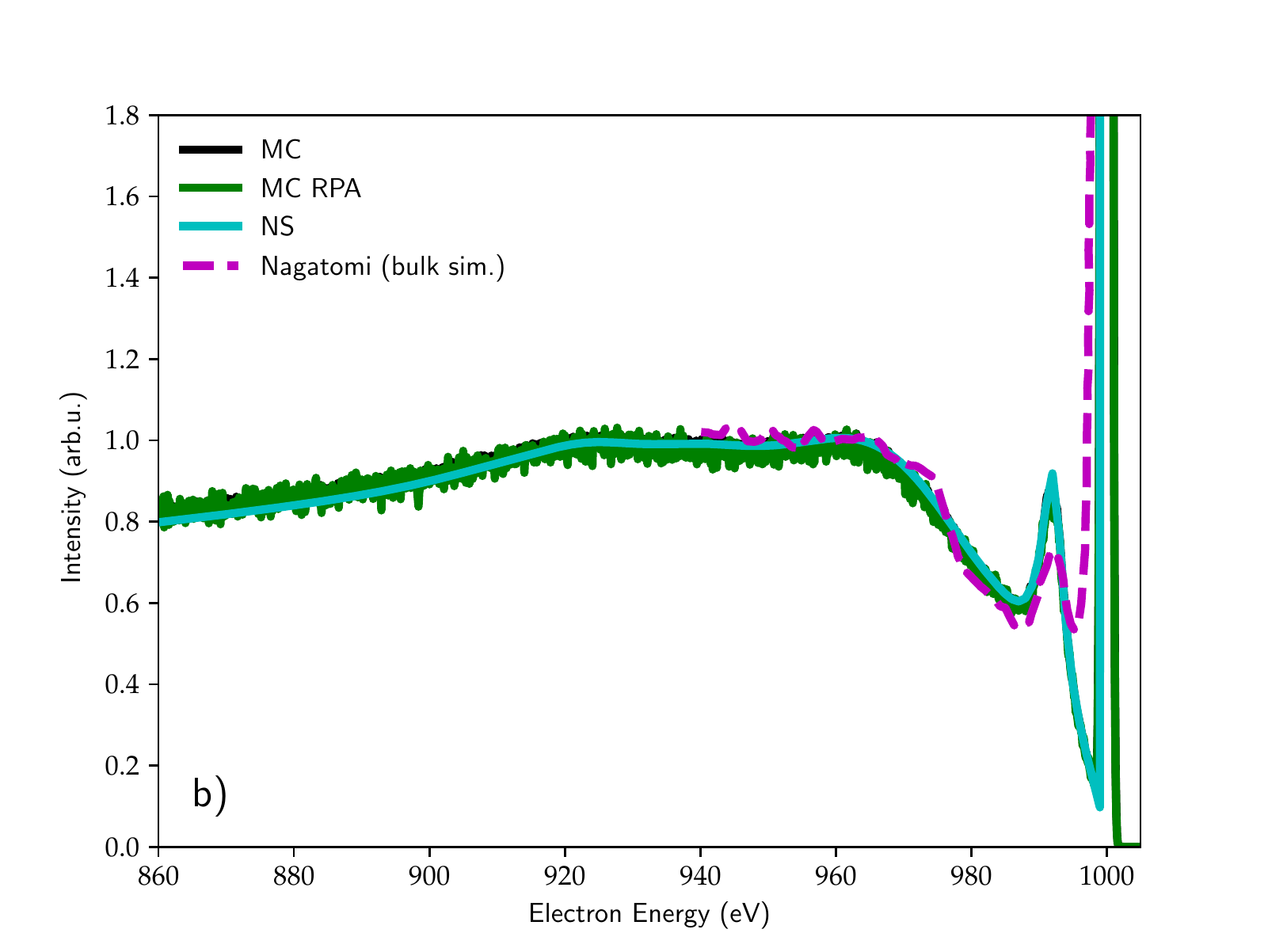}
\includegraphics[width=0.3\linewidth, angle = 0]{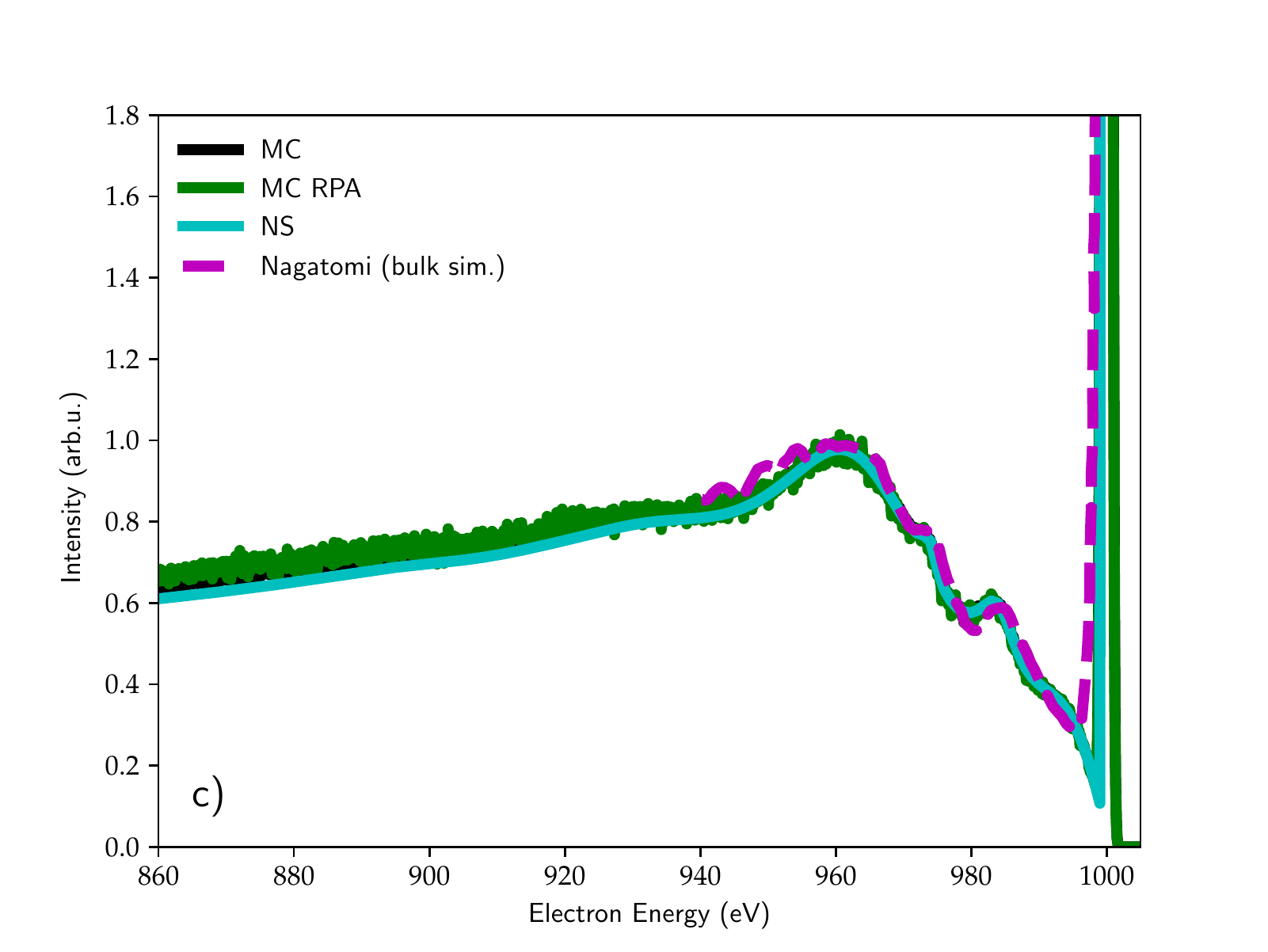}
\caption{REELS of a) Cu, b) Ag, and c) Au calculated using the full 3D bulk (RPA, green curve, see Eq. (\ref{fullRPA})) and high-limit (black curve, see Eq. (\ref{hqlimit})) dispersion laws in comparison to NS (cyan curve) and simulations by Nagatomi et al. (violet curve \cite{nagatomi2003construction}).
 \label{REELbulk}}
\end{figure}

The REEL spectra simulated by the MC and NS approaches using the bulk ELFs are reported in Fig. \ref{REELbulk} a), b), and c) for Cu, Ag and Au, respectively. The spectra are also compared to the simulations carried out by Nagatomi {\it et al.} \citep{nagatomi2003construction} (violet curve) and to the NS approach (cyan curve). 
Our REEL spectra are obtained by using the full bulk ELF (green curve, see Eq. (\ref{fullRPA})) and its quadratic approximation (black curve, see Eq. (\ref{hqlimit})) at high momenta.\\ 
\indent We notice that the agreement between our simulated spectra and those calculated spectra by Nagatomi \citep{nagatomi2003construction} is rather satisfactory, as well as the agreement obtained by using the two different numerical approaches. This means that, for a given set of input data, the accuracy of MC and NS can be considered comparable for all intents and purposes.

\begin{figure*}[h!]
\centering
\includegraphics[width=0.88\textwidth]{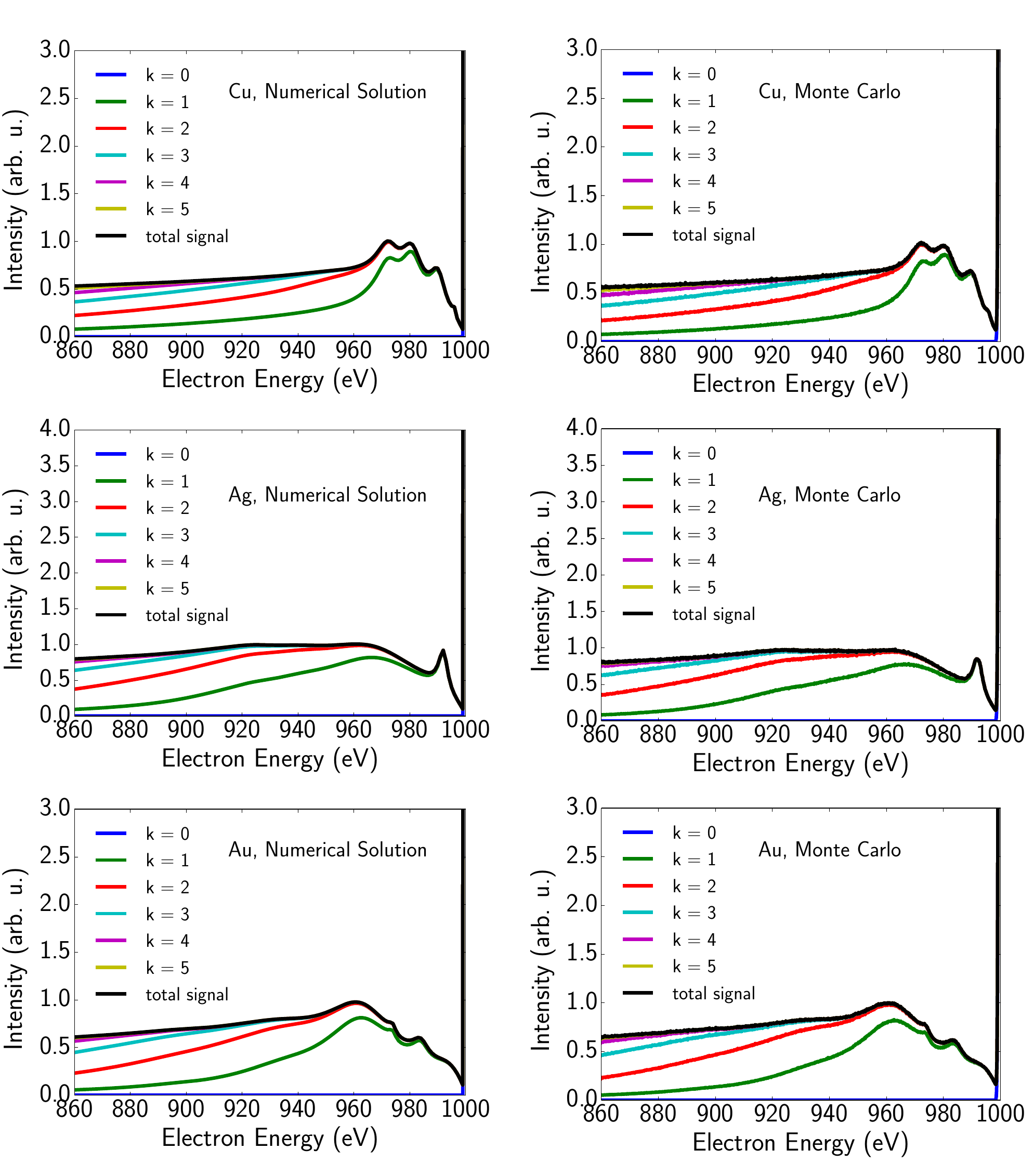}
\caption{REEL spectra of Cu, Ag and Au, deconvoluted for the number of inelastic collisions that electrons undergo in their way out of the solid. On the left panel we show MC simulations, while on the right panel we report the same quantities obtained by the NS approach. The primary beam kinetic energy is set to 1000 eV. The data are normalized at a common height of the main plasmon peak.  \label{fig:reelNin}}
\end{figure*}

In Fig. \ref{fig:reelNin} we report the REEL spectra deconvoluted for the number of inelastic interactions that electrons at most undergo in their way out of the solid. We conclude that the main plasmon peak in the relevant REEL spectra is due to electrons inelastically scattered only once. At variance, electrons experiencing more than one inelastic collision contribute to the REEL intensity at energies beyond the main plasmon peak, leading to the so-called multiple plasmon excitations. Furthermore, we notice the good agreement between the spectral behavior obtained using the two different methods also in this case.
\newpage

\subsection{Effective ELF}

\begin{figure}[h!]
\centering
\includegraphics[width=0.3\linewidth, angle = 0]{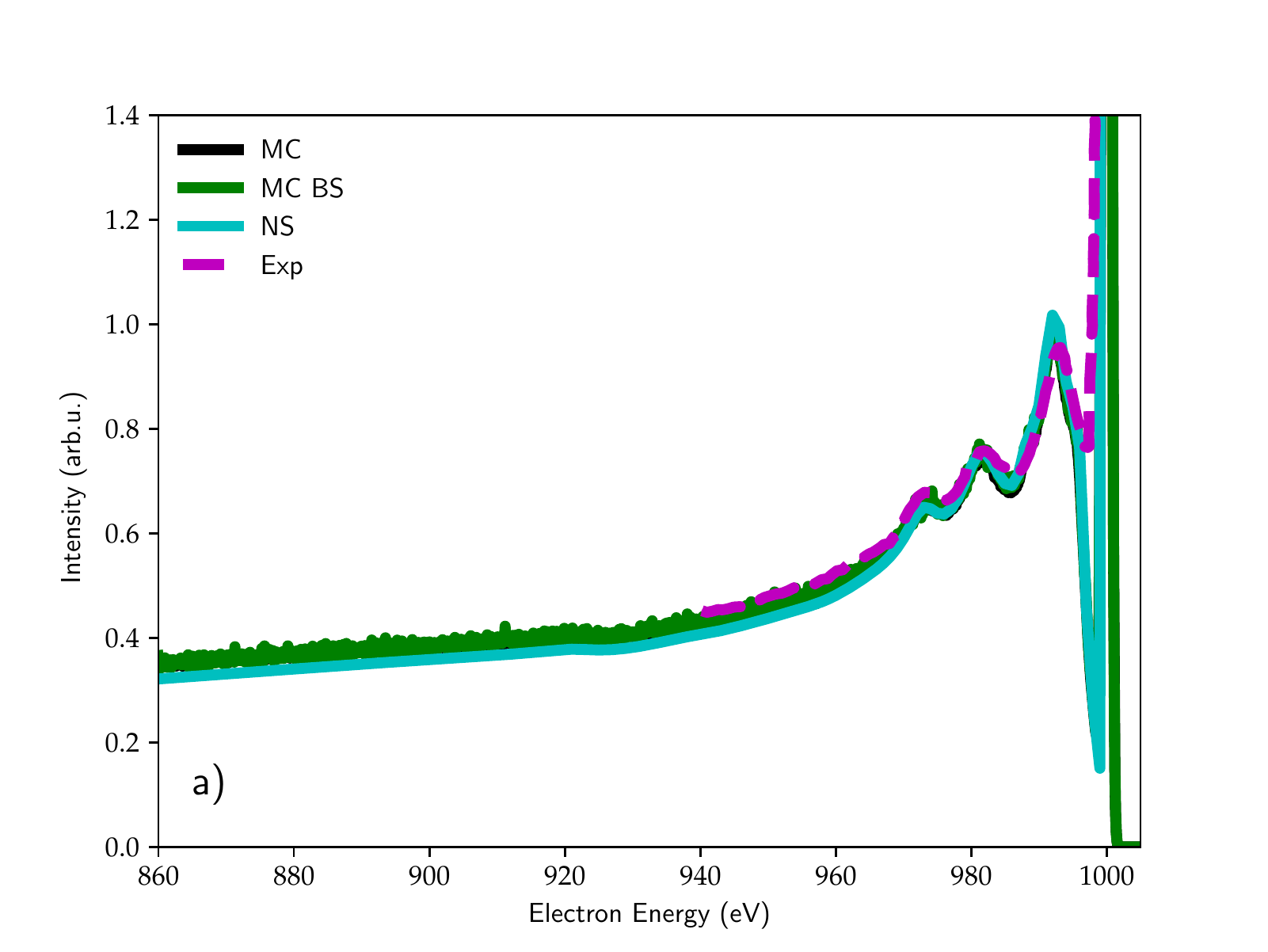}
\includegraphics[width=0.3\linewidth, angle = 0]{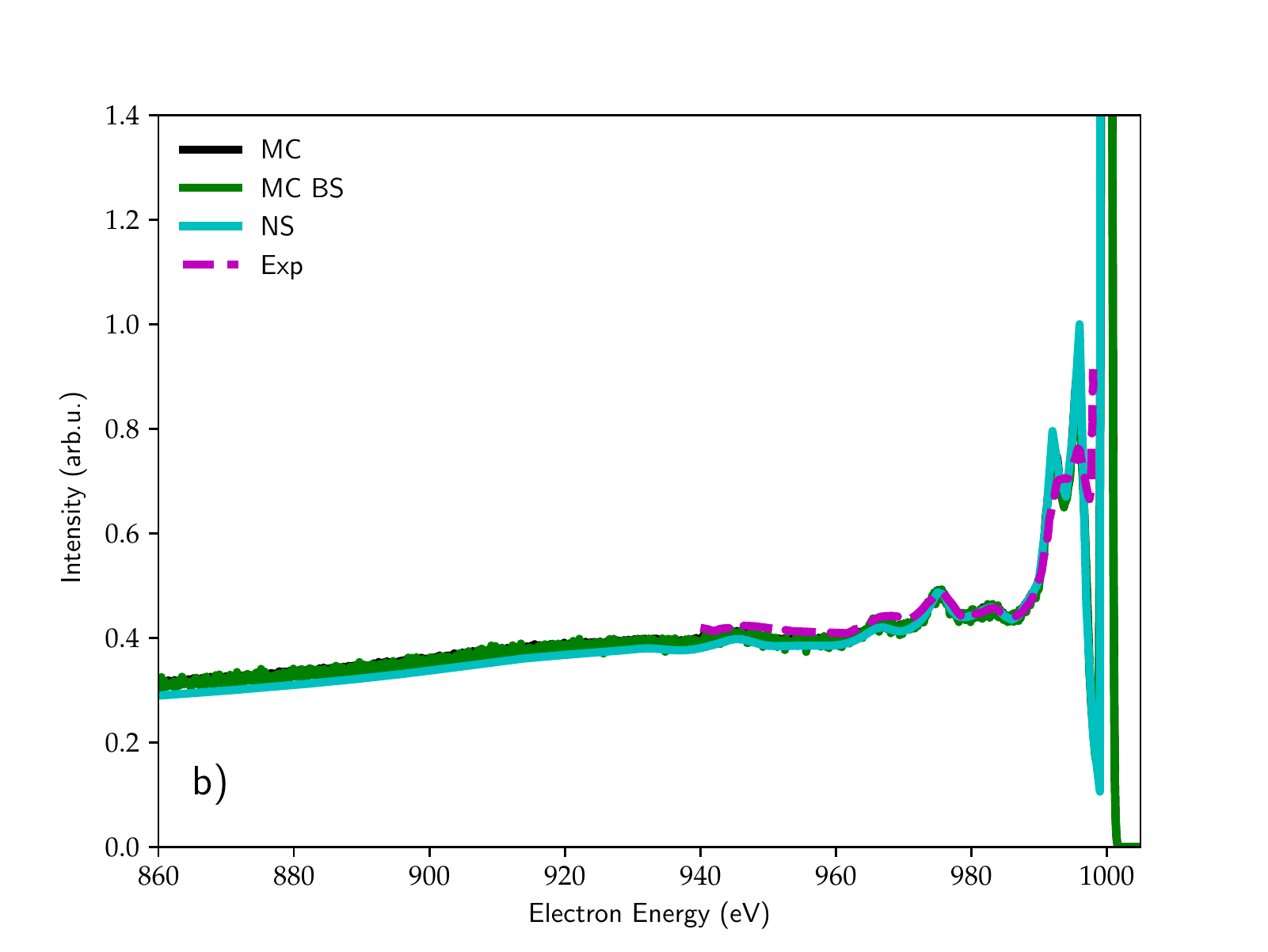}
\includegraphics[width=0.3\linewidth, angle = 0]{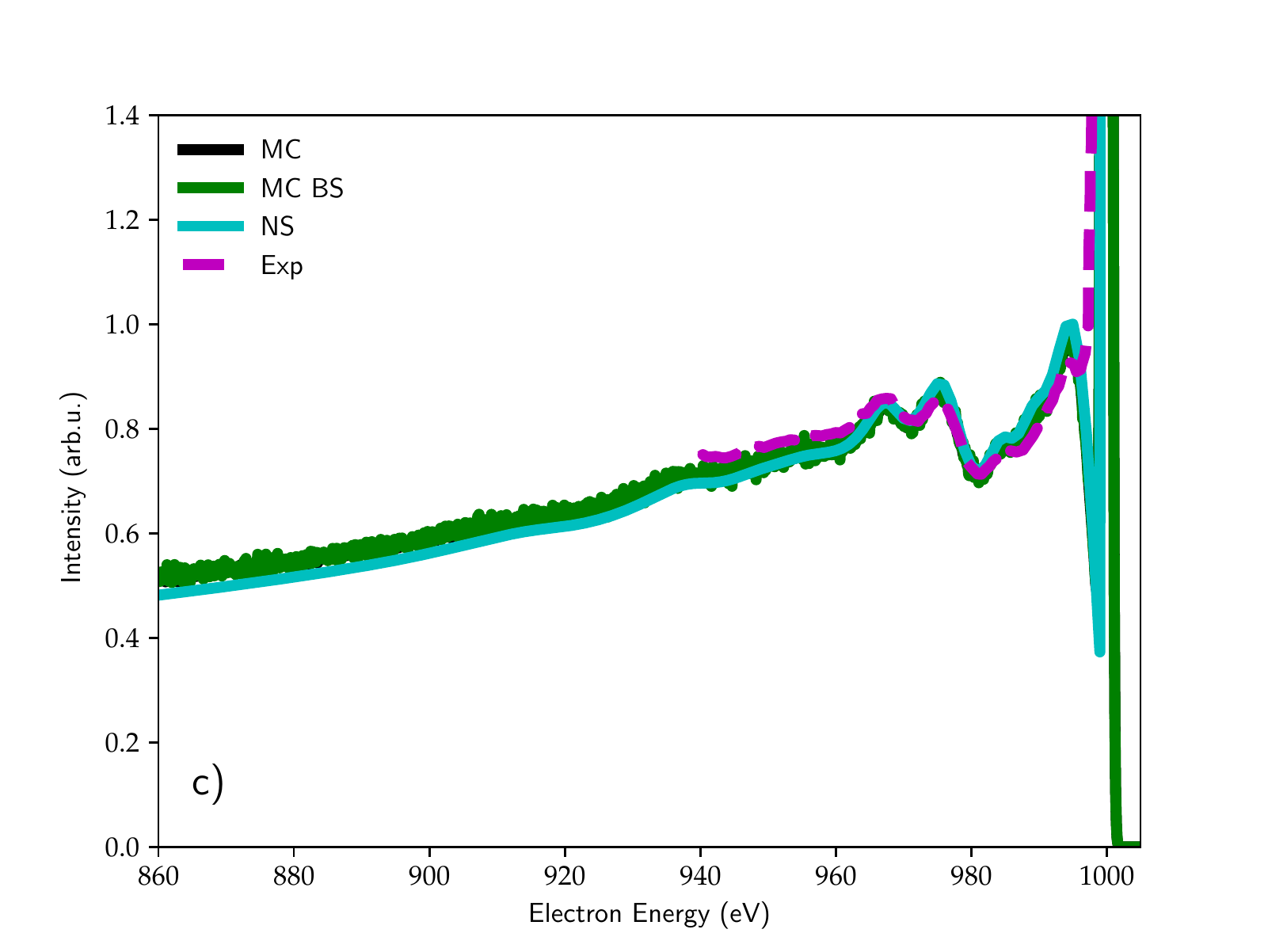}
\caption{REELS of a) Cu, b) Ag, and c) Au calculated using the effective ELF with the dispersion laws of Eq. (\ref{SB}) (BS, green curve) and of Eq. (\ref{fullRPA}) (bulk, black curve) in comparison to NS (cyan curve) and experimental spectra by Nagatomi et al. (violet curve \cite{nagatomi2003construction} ).
 \label{REELBS}}
\end{figure}

Adopting an effective description of the ELFs to include surface plasmon contributions, we calculated the REELs spectra of Cu, Ag and Au by using both the MC and NS approaches. Our simulations are compared with the relevant experimental data \citep{nagatomi2003construction} in Fig. \ref{REELBS}. We notice that the use of effective ELFs, including the contribution of both surface and bulk plasmons, leads to theoretical spectral lineshapes in good agreement with REEL measurements.   
To carry out these simulations we used the effective ELF with the dispersion law of Eq. (\ref{SB}) (BS, green curve) and of Eq. (\ref{fullRPA}) (RPA, black curve) in comparison to NS (cyan curve) and experimental spectra by Nagatomi et al. (violet curve \cite{nagatomi2003construction})

\begin{figure*}[h!]
\centering
\includegraphics[width=0.62\textwidth, angle = 0]{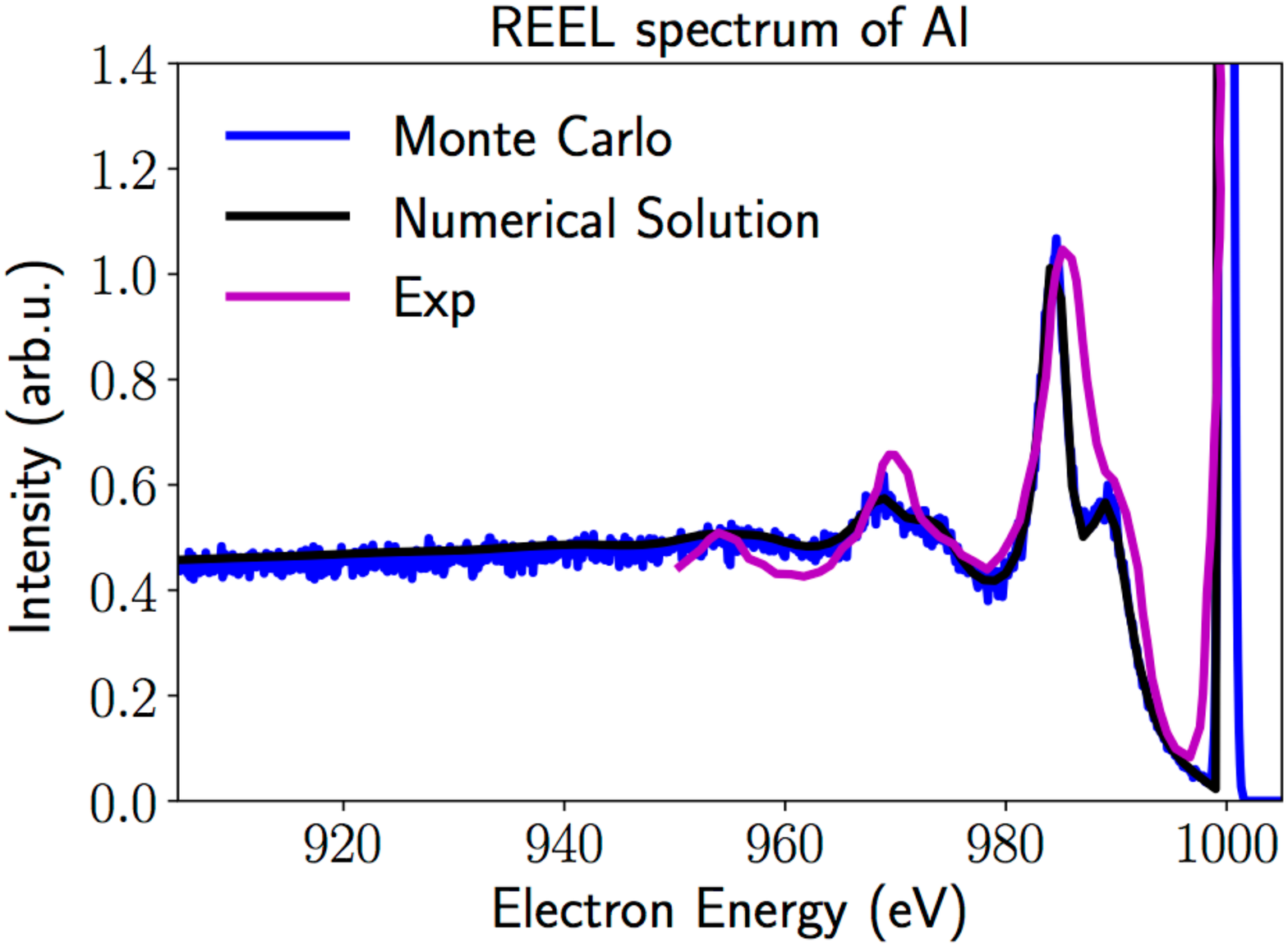}
\caption{REEL spectra of Al obtained by the NS (black curves) and MC approaches, for a primary beam kinetic energy equal to 1000 eV, starting from the Effective ELF. Our results are compared to the experimental spectra by Nagatomi et al \citep{nagatomi2003construction} (magenta lines). The data are normalized in such a way that the signal at 955 eV (inelastic background) is the same in both the theoretical and experimental spectra. \label{fig:reelAl}}
\end{figure*}

The good agreement of the REEL spectra, independently of the dispersion law, basically shows that the latter have negligible effect on the REEL spectra of Cu, Ag, and Au in the selected energy range. 
Thus, even though the momentum dispersion for bulk and surface plasmons has different spatial behaviour \cite{Kyriakou}, on the basis of the analysis of the IMFPs and REEL spectra we conclude that our results can be regarded as robust with respect to momentum transfer dependence in Cu, Ag, and Au in the range of energies of 1 keV to 900 eV.\\
\indent Finally, in order to quantify the respective contribution of bulk and surface inelastic scattering events to the REEL spectra, we simulate the Al EEL spectra, where the surface and bulk plasmon peaks are well-resolved in energy,  by using both MC and NS. The effective energy-loss function of Al was taken from Ref. \citep{nagatomi2003construction}. In Fig. \ref{fig:reelAl} we report the results of our simulations, where we show again an excellent agreement between the two methods and the experimental spectra. Nevertheless, we observe that in the case of our simulations the surface peak is clearly well resolved at odds with  the experimental spectrum owing to the relatively low energy resolution (0.25\% according to Nagatomi {\it et al.} \citep{nagatomi2003construction}). 
In light of these results, we notice that the accuracy achievable by MC and NS approaches in REEL spectra simulations is comparable, once the ELF and its extension beyond the optical limit are carefully assessed. 
\section{Conclusions}

In this work, the performances of MC method against NS of the Ambartsumian-Chandrasekhar equations were compared in terms of accuracy and computational cost for calculating REEL spectra of several metals, such as Cu, Ag and Au. 
To obtain a fair comparison, the computations were performed using the same input data in each test case, which means the same IMFPs for the relevant test cases. 
Indeed, both these approaches are very versatile as they basically require in input information on the materials dielectric response to electromagnetic fields, which can be retrieved by using several tools, from experimental measurements to {\it ab-initio} simulations \citep{azzolini2017monte}.
The spectra obtained with these two methods are comparable and show a good agreement with experimental data, even when including an effective description of the ELF for dealing with surface plasmon excitations. To account for surface excitations, the investigated samples are considered in the NS approach as multi-layer systems, where the surface is characterised by a layer with different scattering properties. \\ 
\indent Using the MC approach one can follow directly the trajectories of the scattered electrons, while within the NS method one solves the integro-differential equations for the partial intensities of scattered electrons with boundary conditions, which are similar to those developed by Ambartsumian and Chandrasekhar to study radiative transfer problems. Basically, the latter method represents indeed a way to solve numerically the transport equation using the backward differential formula, which is equivalent to follow the electron trajectory via a MC algorithm. The NS approach can handle both linear and nonlinear problems and provides numerically exact solutions. 
We notice that the only source of error in the NS approach, other than the V-trajectory approximation,  is due to the integration of the equation in a grid and, as much as in the MC method, can be systematically reduced. applying, the convergence to the physical solution, which depends on the geometry and physical environment of the medium through which the electrons flow, is not automatically guaranteed in the NS approach. \\
\indent It is notable that the NS method is considerably less expensive than MC with respect to computational performance in the simulation of the REEL spectra. For instance, it takes less than 0.5 sec to reproduce the REEL spectrum shown in Fig. \ref{REELbulk} on an Intel Xeon CPU E5-1620 3.60 GHz. More in detail, a comparison of the computational efficiency of these two methods was performed by using 2.9 Ghz Intel Core i7 processor. On the one hand, the NS method running on one processor takes only a total CPU time equal to 10 seconds to output a full REEL spectrum. On the other hand, to deliver the REEL spectrum from MC simulations the use of 4 CPUs for 70 minutes is required. This suggests that the NS technique is a very promising and efficient tool for simulating REEL spectra.\\ 
\indent Nevertheless, while MC calculations require a larger CPU time, they may provide the full spectrum of emitted electrons, which includes multiple scattering and also the emission of secondary electrons. The latter information is crucial e.g. to view images in the scanning electron microscope (SEM) and cannot be achieved at this stage by the NS method. In fact, a shortcoming of the NS approach can be found in its extension to include other scattering mechanisms, such as for example the production of secondary electrons due to direct photoionization, resonant autoionization events, and Auger decay processes \cite{TAIOLI2010237}. The number of electrons is indeed constant in the transport equation. While this issue could be in principle solved by including {\it ad hoc} source and drain terms in some regions, the description of non-equilibrium systems, where electrons form and disappear, in the NS model is still a major challenge. In MC approaches this extension, such as for example for modelling the low-energy harmful secondary electron formation in cancer treatment in biomedical context, has been already successfully pursued \cite{PhysRevB.96.064113}.\\
\indent Finally, we also remind that by using our MC approach is in principle possible to assess the contribution of surface plasmons also starting from a bulk ELF and without using the effective ELF \cite{doi:10.1002/sia.5495}. This is not the case of the NS approach, which cannot retrieve the surface plasmonic characteristics starting from the bulk ELF.

\section*{Data Availability Statement}
The datasets generated for this study can be found in the Mendley repository\\ (http://dx.doi.org/10.17632/d5w45jk95c.1, http://dx.doi.org/10.17632/sv9gzjz5fx.1,\\ http://dx.doi.org/10.17632/4tj26w9w6j.1, http://dx.doi.org/10.17632/s862yyvw5h.1.)

\section*{Acknowledgements}

N.M.P. is supported by the European Commission under the Graphene Flagship Core 2 grant No. 
785219 (WP14, ``Composites''), the FET Proactive (``Neurofibres'') grant No. 732344 as well as by the Italian Ministry of Education, University and Research (MIUR) under the ``Departments of Excellence'' grant L.232/2016, the ARS01-01384-PROSCAN and the PRIN-20177TTP3S grants. The authors gratefully acknowledge the Gauss Centre for Supercomputing for funding this project by providing computing time on the GCS Supercomputer JUQUEEN at J{\"u}lich Supercomputing Centre (JSC) \cite{juqueen}. Furthermore, the authors acknowledge Bruno Kessler Foundation (FBK) for providing unlimited access to the KORE computing facility. 

%\FloatBarrier

\section*{References}
%\bibliographystyle{unsrt.bst}
%\bibliography{Biblio.bib}

\begin{thebibliography}{10}

\bibitem{hillier1944microanalysis}
J~Hillier and RF~Baker.
\newblock Microanalysis by means of electrons.
\newblock {\em Journal of Applied Physics}, 15(9):663, 1944.

\bibitem{yubero1998optical}
F~Yubero, VM~Jim{\'e}nez, and AR~Gonz{\'a}lez-Elipe.
\newblock Optical properties and electronic transitions of {SnO$_2$} thin films
  by reflection electron energy loss spectroscopy.
\newblock {\em Surface science}, 400(1):116, 1998.

\bibitem{nikzad1992quantitative}
S~Nikzad, CC~Ahn, and HA~Atwater.
\newblock Quantitative analysis of semiconductor alloy composition during
  growth by reflection-electron energy loss spectroscopy.
\newblock {\em Journal of Vacuum Science \& Technology B: Microelectronics and
  Nanometer Structures Processing, Measurement, and Phenomena}, 10(2):762,
  1992.

\bibitem{da2014monte}
B~Da, ZY~Li, HC~Chang, SF~Mao, and ZJ~Ding.
\newblock A monte carlo study of reflection electron energy loss spectroscopy
  spectrum of a carbon contaminated surface.
\newblock {\em Journal of Applied Physics}, 116(12):124307, 2014.

\bibitem{RitchieHowie}
RH~Ritchie and A~Howie.
\newblock Electron excitation and the optical potential in electron microscopy.
\newblock {\em The Philosophical Magazine: A Journal of Theoretical
  Experimental and Applied Physics}, 36(2):463, 1977.

\bibitem{egerton2011electron}
RF~Egerton.
\newblock {\em Electron energy-loss spectroscopy in the electron microscope}.
\newblock Springer Science \& Business Media, 2011.

\bibitem{TAIOLI2015191}
Simone Taioli and Stefano Simonucci.
\newblock Chapter five - a computational perspective on multichannel scattering
  theory with applications to physical and nuclear chemistry.
\newblock volume~11 of {\em Annual Reports in Computational Chemistry}, page
  191. Elsevier, 2015.

\bibitem{PhysRevB.79.085432}
Simone Taioli, Stefano Simonucci, Lucia Calliari, Massimiliano Filippi, and
  Maurizio Dapor.
\newblock Mixed ab initio quantum mechanical and monte carlo calculations of
  secondary emission from {SiO$_{2}$} nanoclusters.
\newblock {\em Phys. Rev. B}, 79:085432, Feb 2009.

\bibitem{doi:10.1002/sia.5947}
Rafael Garcia-Molina, Isabel Abril, Ioanna Kyriakou, and Dimitris Emfietzoglou.
\newblock Inelastic scattering and energy loss of swift electron beams in
  biologically relevant materials.
\newblock {\em Surface and Interface Analysis}, 49(1):11--17, 2017.

\bibitem{doi:10.1002/sia.5878}
Dimitris Emfietzoglou, Ioanna Kyriakou, Rafael Garcia-Molina, and Isabel Abril.
\newblock Inelastic mean free path of low-energy electrons in condensed media:
  beyond the standard models.
\newblock {\em Surface and Interface Analysis}, 49(1):4--10, 2017.

\bibitem{umari2012communication}
P~Umari, O~Petrenko, Simone Taioli, and MM~De~Souza.
\newblock Communication: electronic band gaps of semiconducting zig-zag carbon
  nanotubes from many-body perturbation theory calculations, 2012.

\bibitem{taioli2009electronic}
S~Taioli, P~Umari, and MM~De~Souza.
\newblock Electronic properties of extended graphene nanomaterials from gw
  calculations.
\newblock {\em physica status solidi (b)}, 246(11-12):2572--2576, 2009.

\bibitem{azzolini2017monte}
M~Azzolini, T~Morresi, G~Garberoglio, L~Calliari, N~M Pugno, S~Taioli, and
  M~Dapor.
\newblock Monte carlo simulations of measured electron energy-loss spectra of
  diamond and graphite: Role of dielectric-response models.
\newblock {\em Carbon}, 118:299, 2017.

\bibitem{Kyriakou}
Ioanna Kyriakou, Dimitris Emfietzoglou, Rafael Garcia-Molina, Isabel Abril, and
  Kostas Kostarelos.
\newblock Simple model of bulk and surface excitation effects to inelastic
  scattering in low-energy electron beam irradiation of multi-walled carbon
  nanotubes.
\newblock {\em Journal of Applied Physics}, 110:054304--054304, 09 2011.

\bibitem{segatta2017quantum}
Francesco Segatta, Lorenzo Cupellini, Sandro Jurinovich, Shaul Mukamel,
  Maurizio Dapor, Simone Taioli, Marco Garavelli, and Benedetta Mennucci.
\newblock A quantum chemical interpretation of two-dimensional electronic
  spectroscopy of light-harvesting complexes.
\newblock {\em Journal of the American Chemical Society}, 139(22):7558, 2017.

\bibitem{azzolini2018secondary}
Martina Azzolini, Marco Angelucci, Roberto Cimino, Rosanna Larciprete, Nicola~M
  Pugno, Simone Taioli, and Maurizio Dapor.
\newblock Secondary electron emission and yield spectra of metals from monte
  carlo simulations and experiments.
\newblock {\em Journal of Physics: Condensed Matter}, 31(5):055901, 2018.

\bibitem{azzolini2018anisotropic}
Martina Azzolini, Tommaso Morresi, Kerry Abrams, Robert Masters, Nicola
  Stehling, Cornelia Rodenburg, Nicola~M Pugno, Simone Taioli, and Maurizio
  Dapor.
\newblock Anisotropic approach for simulating electron transport in layered
  materials: Computational and experimental study of highly oriented pyrolitic
  graphite.
\newblock {\em The Journal of Physical Chemistry C}, 122(18):10159--10166,
  2018.

\bibitem{Lindhard}
J.~Lindhard and K.~Dan. Vidensk.
\newblock {\em Selsk. Mat. Fys. Medd.}, 1954.

\bibitem{PhysRevB.35.482}
David~R. Penn.
\newblock Electron mean-free-path calculations using a model dielectric
  function.
\newblock {\em Phys. Rev. B}, 35:482--486, Jan 1987.

\bibitem{PhysRevB.1.2362}
N.~D. Mermin.
\newblock Lindhard dielectric function in the relaxation-time approximation.
\newblock {\em Phys. Rev. B}, 1:2362--2363, Mar 1970.

\bibitem{dapor2014transport}
M~Dapor.
\newblock {\em Transport of energetic electrons in solids}, volume 257.
\newblock Springer, 2017.

\bibitem{1749-4699-2-1-015002}
Simone Taioli, Stefano Simonucci, and Maurizio Dapor.
\newblock Surprises: when ab initio meets statistics in extended systems.
\newblock {\em Computational Science \& Discovery}, 2(1):015002, 2009.

\bibitem{afanas2016analytical}
VP~Afanasev, DS~Efremenko, and PS~Kaplya.
\newblock Analytical and numerical methods for computing electron partial
  intensities in the case of multilayer systems.
\newblock {\em Journal of Electron Spectroscopy and Related Phenomena}, 210:16,
  2016.

\bibitem{mott1929scattering}
NF~Mott.
\newblock The scattering of fast electrons by atomic nuclei.
\newblock {\em Proceedings of the Royal Society of London. Series A, Containing
  Papers of a Mathematical and Physical Character}, 124(794):425, 1929.

\bibitem{Ritchie_PhysRev_1957}
RH~Ritchie.
\newblock Plasma losses by fast electrons in thin films.
\newblock {\em Phys. Rev.}, 106:874, Jun 1957.

\bibitem{database_input}
M~Azzolini.
\newblock Input data set of {C}u, {A}g and {A}u, regarding electron elastic and
  inelastic scattering, for the calculation of the reflection electron energy
  loss spectra.
\newblock Mendeley Data, v1, 2018.

\bibitem{inelastic_calc}
M~Dapor and M~Azzolini.
\newblock {R}itchie dielectric theory: calculation of the inelastic mean free
  path, of the differential inverse inelastic mean free path and of the
  cumulative inelastic probability distribution.
\newblock Mendeley Data, v1, 2018.

\bibitem{REEL_NS_calc}
P~Kaplya, O.~Yu Ridzel, and M~Azzolini.
\newblock Calculation of {REEL} spectra by applying the numerical solution of
  the {A}mbartsymian-{C}handrasekhar equation.
\newblock Mendeley Data, v1, 2018.

\bibitem{elastic_calc}
P~Kaplya, O.~Yu Ridzel, and M~Azzolini.
\newblock Electron elastic scattering treated with the {M}ott theory:
  calculation of the elastic mean free path, of the differential elastic cross
  section and of the cumulative elastic probability distribution.
\newblock Mendeley Data, v1, 2018.

\bibitem{doi:10.1063/1.5000118}
L.~A. Gonzalez, M.~Angelucci, R.~Larciprete, and R.~Cimino.
\newblock The secondary electron yield of noble metal surfaces.
\newblock {\em AIP Advances}, 7(11):115203, 2017.

\bibitem{TAIOLI2010237}
Simone Taioli, Stefano Simonucci, Lucia Calliari, and Maurizio Dapor.
\newblock Electron spectroscopies and inelastic processes in nanoclusters and
  solids: Theory and experiment.
\newblock {\em Physics Reports}, 493(5):237, 2010.

\bibitem{chandras}
S.~Chandrasekhar.
\newblock {\em Radiative Transfer}.
\newblock Dover Books on Intermediate and Advanced Mathematics. Dover
  Publications, 1960.

\bibitem{Afanas2017}
Viktor~P Afanas, Alexander~S Gryazev, Dmitry~S Efremenko, and Pavel~S Kaplya.
\newblock {Differential inverse inelastic mean free path and differential
  surface excitation probability retrieval from electron energy loss spectra}.
\newblock {\em Vaccum}, 136, 2017.

\bibitem{Werner2001}
Wolfgang S.~M. Werner.
\newblock {Electron transport in solids for quantitative surface analysis}.
\newblock {\em Surface and Interface Analysis}, 31(3):141--176, mar 2001.

\bibitem{Salvat2005}
Francesc Salvat, Aleksander Jablonski, and Cedric~J. Powell.
\newblock {ELSEPA — Dirac partial-wave calculation of elastic scattering of
  electrons and positrons by atoms, positive ions and molecules}.
\newblock {\em Computer Physics Communications}, 165(2):157, jan 2005.

\bibitem{Afanasev1994}
V.~P. Afanas'ev, S.~D. Fedorovich, A.~V. Lubenchenko, A.~A. Ryjov, and M.~S.
  Esimov.
\newblock {Kilovolt electron backscattering}.
\newblock {\em Zeitschrift f{\"{u}}r Physik B Condensed Matter}, 96(2):253,
  1994.

\bibitem{Afanasev2015}
V~P {Afanas 'ev}, O~Y Golovina, A~S Gryazev, D~S Efremenko, and P~S Kaplya.
\newblock {Photoelectron spectra of finite-thickness layers}.
\newblock {\em Journal of Vacuum Science and Technology B}, 33:3, 2015.

\bibitem{salvat1987analytical}
F~Salvat, JD~Martnez, R~Mayol, and J~Parellada.
\newblock Analytical dirac-hartree-fock-slater screening function for atoms
  ({Z}= 1--92).
\newblock {\em Physical Review A}, 36(2):467, 1987.

\bibitem{jablonski2004comparison}
A~Jablonski, F~Salvat, and CJ~Powell.
\newblock Comparison of electron elastic-scattering cross sections calculated
  from two commonly used atomic potentials.
\newblock {\em Journal of physical and chemical reference data}, 33(2):409,
  2004.

\bibitem{salvat1993elastic}
F~Salvat and R~Mayol.
\newblock Elastic scattering of electrons and positrons by atoms.
  {S}chr{\"o}dinger and {D}irac partial wave analysis.
\newblock {\em Computer physics communications}, 74(3):358, 1993.

\bibitem{dapor1996elastic}
M~Dapor.
\newblock Elastic scattering calculations for electrons and positrons in solid
  targets.
\newblock {\em Journal of applied physics}, 79(11):8406, 1996.

\bibitem{salvat2005elsepa}
F~Salvat, A~Jablonski, and CJ~Powell.
\newblock Elsepa—dirac partial-wave calculation of elastic scattering of
  electrons and positrons by atoms, positive ions and molecules.
\newblock {\em Computer physics communications}, 165(2):157, 2005.

\bibitem{tung1994differential}
CJ~Tung, YF~Chen, CM~Kwei, and TL~Chou.
\newblock Differential cross sections for plasmon excitations and reflected
  electron-energy-loss spectra.
\newblock {\em Physical Review B}, 49(23):16684, 1994.

\bibitem{CHEN1996131}
Y.F. Chen and C.M. Kwei.
\newblock Electron differential inverse mean free path for surface electron
  spectroscopy.
\newblock {\em Surface Science}, 364(2):131 -- 140, 1996.

\bibitem{doi:10.1002/sia.5495}
Lucia Calliari, Maurizio Dapor, Giovanni Garberoglio, and Sergey Fanchenko.
\newblock Momentum transfer dependence of reflection electron energy loss
  spectra: theory and experiment.
\newblock {\em Surface and Interface Analysis}, 46(5):340--349, 2014.

\bibitem{10.1667/RR3281}
Dimitris Emfietzoglou and Hooshang Nikjoo.
\newblock {The Effect of Model Approximations on Single-Collision Distributions
  of Low-Energy Electrons in Liquid Water}.
\newblock {\em Radiation Research}, 163(1):98 -- 111, 2005.

\bibitem{doi:10.1063/1.4824541}
Dimitris Emfietzoglou, Ioanna Kyriakou, Rafael Garcia-Molina, and Isabel Abril.
\newblock The effect of static many-body local-field corrections to inelastic
  electron scattering in condensed media.
\newblock {\em Journal of Applied Physics}, 114(14):144907, 2013.

\bibitem{tanuma2011calculations}
S~Tanuma, CJ~Powell, and DR~Penn.
\newblock Calculations of electron inelastic mean free paths. {IX}. data for 41
  elemental solids over the 50 e{V} to 30 ke{V} range.
\newblock {\em Surface and Interface Analysis}, 43(3):689, 2011.

\bibitem{denton2008influence}
CD~Denton, I~Abril, JC~Garcia-Molina, Rand Moreno-Mar{\'\i}n, and
  S~Heredia-Avalos.
\newblock Influence of the description of the target energy-loss function on
  the energy loss of swift projectiles.
\newblock {\em Surface and Interface Analysis}, 40(11):1481, 2008.

\bibitem{montanari2007calculation}
CC~Montanari, JE~Miraglia, S~Heredia-Avalos, R~Garcia-Molina, and I~Abril.
\newblock Calculation of energy-loss straggling of {C}, {Al}, {Si}, and {Cu}for
  fast {H}, {He}, and {L}i ions.
\newblock {\em Physical Review A}, 75(2):022903, 2007.

\bibitem{smith1985handbook}
DY~Smith, E~Shiles, M~Inokuti, and ED~Palik.
\newblock Handbook of optical constants of solids.
\newblock {\em Handbook of Optical Constants of Solids}, 1:369, 1985.

\bibitem{HENKE1993181}
B.L. Henke, E.M. Gullikson, and J.C. Davis.
\newblock X-ray interactions: Photoabsorption, scattering, transmission, and
  reflection at e = 50-30,000 ev, z = 1-92.
\newblock {\em Atomic Data and Nuclear Data Tables}, 54(2):181 -- 342, 1993.

\bibitem{doi:10.1002/sia.740111107}
S.~Tanuma, C.~J. Powell, and D.~R. Penn.
\newblock Calculations of electron inelastic mean free paths for 31 materials.
\newblock {\em Surface and Interface Analysis}, 11(11):577--589, 1988.

\bibitem{nagatomi2003construction}
T~Nagatomi, Y~Takai, BV~Crist, K~Goto, and R~Shimizu.
\newblock Construction of database of effective energy-loss functions.
\newblock {\em Surface and interface analysis}, 35(2):174, 2003.

\bibitem{yoshikawa1995h}
H~Yoshikawa.
\newblock H. yoshikawa, y. irokawa, and r. shimizu, j. vac. sci. technol. a 13,
  1984 (1995).
\newblock {\em Journal of Vacuum Science and Technology A}, 13:1984, 1995.

\bibitem{PhysRevB.96.064113}
Maurizio Dapor, Isabel Abril, Pablo de~Vera, and Rafael Garcia-Molina.
\newblock Energy deposition around swift proton tracks in
  polymethylmethacrylate: How much and how far.
\newblock {\em Phys. Rev. B}, 96:064113, Aug 2017.

\bibitem{juqueen}
{J\"{u}lich Supercomputing Centre}.
\newblock {JUQUEEN: IBM Blue Gene/Q Supercomputer System at the J\"{u}lich
  Supercomputing Centre}.
\newblock {\em Journal of large-scale research facilities}, 1(A1), 2015.

\end{thebibliography}

\end{document}